\documentclass[11pt,english,onecolumn,draftcls]{IEEEtran}
\usepackage[T1]{fontenc}
\usepackage[latin9]{inputenc}
\usepackage[active]{srcltx}
\usepackage{color}
\usepackage{float}
\usepackage{amsthm}
\usepackage{amssymb}
\usepackage{graphicx}
\usepackage{epstopdf}

\makeatletter

\providecommand{\tabularnewline}{\\}
\floatstyle{ruled}
\newfloat{algorithm}{tbp}{loa}
\providecommand{\algorithmname}{Algorithm}
\floatname{algorithm}{\protect\algorithmname}

\theoremstyle{plain}
\newtheorem{thm}{\protect\theoremname}
\theoremstyle{plain}
\newtheorem{prop}[thm]{\protect\propositionname}
\theoremstyle{plain}
\newtheorem{lem}[thm]{\protect\lemmaname}
\theoremstyle{plain}
\newtheorem{cor}[thm]{\protect\corollaryname}

\@ifundefined{showcaptionsetup}{}{%
 \PassOptionsToPackage{caption=false}{subfig}}
\usepackage{subfig}
\makeatother

\usepackage{babel}
\providecommand{\corollaryname}{Corollary}
\providecommand{\lemmaname}{Lemma}
\providecommand{\propositionname}{Proposition}
\providecommand{\theoremname}{Theorem}

\begin{document}

\title{Efficient Approximation Algorithms for Multi-Antennae Largest Weight
Data Retrieval%
\thanks{The paper has been accepted by IEEE Transactions on Mobile Computing.
Citation information: DOI 10.1109/TMC.2017.2696009. %
}}

\author{\textcolor{black}{Longkun Guo$^{1}$, Hong Shen$^{2,\,3}$, Wenxing
Zhu$^{1}$}\linebreak{}
\textcolor{black}{{} $^{1}$College of Mathematics and Computer Science,
Fuzhou University, China}\\
\textcolor{black}{{} $^{2}$School of Computer Science, University of
Adelaide, Australia }\\
\textcolor{black}{$^{3}$School of Information Science and Technology,
Sun Yat-Sen University, China}\\
}
\maketitle
\begin{abstract}
In a mobile network, wireless data broadcast over $m$ channels (frequencies)
is a powerful means for distributed dissemination of data to clients
who access the channels through multi-antennae equipped on their mobile
devices. The $\delta$-antennae largest weight data retrieval ($\delta$ALWDR)
problem is to compute a schedule for downloading a subset of data
items that has a maximum total weight using $\delta$ antennae in
a given time interval. In this paper, we first give a linear programming
(LP) relaxation for $\delta$ALWDR and show that it is polynomial-time
solvable when every data item appears at most once. We also show that
when there exist data items with multiple occurrences, the integrality
gap of this LP formula is $2$.

We then present an approximation algorithm of ratio $1-\frac{1}{e}$
for the $\delta$-antennae $\gamma$-separated largest weight data
retrieval ($\delta$A$\gamma$LWDR) problem, a weaker version of $\delta$ALWDR
where each block of up to $\gamma$ data (time) slots is separated
by a vacant slot on all channels, applying the techniques called collectively
randomized LP rounding and layered DAG construction. We show that
$\delta$A$\gamma$LWDR is ${\cal NP}$-complete even for the simple
case of $\gamma=2$, $m=3$, and equal-weight data items each appearing
up to 3 times. Our algorithm runs in time $O(2^{\gamma}m^{7}T^{3.5}L)$,
where $T$ is the number of time slots, and $L$ is the maximum length
of the input. Then, from the simple observation that a ratio $\alpha$
approximation solution to $\delta$A$\gamma$LWDR implies a ratio
$\alpha-\epsilon$ approximation solution to $\delta$ALWDR for any
fixed $\epsilon>0$, we immediately have an approximation algorithm
of ratio $1-\frac{1}{e}-\epsilon$ for $\delta$ALWDR. Our algorithm
has the same approximation ratio as the known result in \cite{lu2014data}
which holds only for $\delta=1$ , with a significantly lower time
complexity of $O(2^{\frac{1}{\epsilon}}\frac{1}{\epsilon}m^{7}T^{3.5}L)$
(improved from $O(\epsilon^{3.5}m^{\frac{3.5}{\epsilon}}T^{3.5}L)$
of \cite{lu2014data}). As a by-product, we also give a fixed-parameter
tractable (fpt-)algorithm of time complexity $O(2^{B}m^{7}T^{3.5}L)$
for $\delta$ALWDR, where $B$ is the number of time slots that contain
data items with multiple occurrences. \end{abstract}


\section{Introductions}

In recent years, wireless data broadcast has been gradually considered
as an attractive data propagation scheme for transferring public information
to a large number of specified mobile devices, in applications ranging
from satellite communications to wireless mobile ad hoc networks.
Most wireless broadcast is between base stations and battery-limited
mobile devices, where a base station emits public information (such
as stock marketing, weather, and traffic) via a number of parallel
channels, and mobile devices within a limited area, using an antenna
(or multiple antennae), listen to the channels and obtain required
data packages. The base stations can coordinate information propagation
to cover a larger scope; Within the covered scope, the mobile client
can move freely among different areas while keeping listening to the
channels for downloading the required data packages; An antenna, equipped
in the mobile client, can only listen to a channel at one time, but
it can switch between the channels by adjusting its frequency.

Wireless data broadcast has already shown its advantages in wireless
networks: Possible power saving, throughput improvement, the communication
efficiency that every transmission by a base station can be received
by all nodes which lie within its communication range, and so on.
At the same time, it brings a number of challenges for data propagation
technologies, which have become popular research topics in recent
years, such as indexing technique, data scheduling, and data retrieval.
Indexing technique and data scheduling are mainly based on the server
side. The former topic investigates the structure of the indexing
information that is emitted by the server to boost clients on finding
the locations of requested data items among the channels. The latter
investigates how the server allocates the data items in proper channels
and at proper time slots, such that clients can quickly accomplish
download tasks. Differently, data retrieval is based on the client
side. The goal is to find a data retrieval sequence retrieving all
requested data items among the channels such that the total access
latency is minimized, where the access latency is the length of the
period from the starting time when the client knows the offset of
each requested data item (by index techniques) to the ending time
when the client downloads all the requested data items.

\subsection{Problem Statement}

Let $D=\{d_{1},\, d_{2},\,\dots,\, d_{n}\}$ be a set of data items
broadcast in channels $c_{1},\, c_{2},\,\dots,\, c_{m}$ in a given
time interval, which is separated into time slots $t_{1},\, t_{2},\,\dots,\, t_{T}$.
Let the data items $d_{1},\, d_{2},\,\dots,\, d_{n}$ be with weights
$w_{1},\dots,w_{n}$, respectively. The largest weight data retrieval
(LWDR) problem is to schedule to download the data items of $D$,
such that the weight of the downloaded items will be maximized \cite{Infocom12LuEfficient}.
When the items are with the same weight, LWDR reduces to the Largest
Number Data Retrieval (LNDR) problem, which is to maximize the number
of the downloaded items in the given time interval. This paper considers
$\delta$ALWDR, i.e., LWDR for mobile devices with $\delta$ antennae.
We say there exist conflicts between two data items iff it is impossible
to retrieve both of them in the same broadcast cycle using a same
antenna. There exist two well-known conflicts for data retrieval problems
(including $\delta$ALWDR): (1) two requested data items at two same
time slots; (2) two adjacent time slots of different channels. The
first conflict is because one antenna can retrieve one channel in
one time slot, while the second is because that the antenna switching
between different channels takes time, typically one time slot.

This paper develops a ratio $1-\frac{1}{e}-\epsilon$ approximation
algorithm for $\delta$ALWDR with both of the two conflicts for any
fixed $\epsilon>0$. To do this, we propose the $\delta$A$\gamma$-separated
LWDR ($\delta$A$\gamma$LWDR) problem, a weaker version of $\delta$ALWDR
which has a vacant time slot in every $\gamma$ time slots. A \textbf{vacant}
\textbf{time slot} is a time slot in which no data item is broadcast.
The reason of developing approximation algorithms for $\delta$A$\gamma$LWDR
instead of $\delta$ALWDR is because any approximation algorithm for
$\delta$A$\gamma$LWDR can be immediately adopted to solve $\delta$ALWDR
with only a small loss in the approximation ratio, as the simple observation
in the following (proof in appendix):
\begin{prop}
\label{prop:simpobsbetw-gamma}If $\delta$A$\gamma$LWDR admits a
ratio $\alpha$ approximation algorithm with runtime $t_{\delta A\gamma LWDR}$,
then for any $\epsilon>0$, $\delta$ALWDR admits a ratio $(\alpha-\epsilon)$
approximation algorithm with runtime $O(\frac{1}{\epsilon})\cdot t_{\delta A\gamma LWDR}$.
\end{prop}
A \textbf{segment} of $\delta$A$\gamma$LWDR is the set of time slots
between two neighbor vacant time slots. Through this paper, we assume
that the number of the segments is $N$. This paper also investigates
$\delta$A$\gamma$LWDR under the \textbf{occurrence assumption} to
better develop the approximation algorithm. The \textbf{occurrence
assumption} is: Each item is broadcast at most once in each segment
in $\delta$A$\gamma$LWDR.

\subsection{Related Works}

For LWDR, i.e. $\delta$ALWDR with $\delta=1$, existing literature
has discussed data retrieval in depth in the client side of wireless
data broadcast. The problem is firstly studied under the assumption
that each client is equipped with one antenna, and is to download
multiple requested data items for one single request. Three heuristic
schemes have been proposed in \cite{hurson2006power} to compute a
data retrieval schedule with minimum times of switching between the
channels. Later, two algorithms have been proposed in \cite{Shi2010efficient}
to extend data retrieval technique to the case that clients are equipped
with multiple antennae. Considering neither of the two conflicts,
paper \cite{Gao2011} gives an algorithm to find a data retrieval
schedule with minimum access latency and with the times of switching
between the channels bounded by a given number. A parameterized heuristic
scheme has been proposed in \cite{Infocom12LuEfficient}, attempting
to solve the minimum cost data retrieval problem and to find a data
retrieval schedule with minimized energy consumption. A factor$-0.5$
approximation algorithm for LWDR also has been presented in the same
paper. The key idea of the approximation is first to convert the relationship
between the broadcast data items and the time slots to a bipartite
graph, and then obtain an approximation solution for LWDR via maximum
matching in the bipartite graph. The time complexity is $O(m^{3}T^{3})$
if employing the Hungarian algorithm \cite{korte2002combinatorial}.
The ratio is then improved to $(1-\frac{1}{e}-\epsilon)$ for any
constant $\epsilon>0$, based on a combination of both linear and
nonlinear programming technique in the algorithm of \cite{lu2014data},
with a significantly increased time complexity of $O(\epsilon^{3.5}m^{\frac{3.5}{\epsilon}}T^{3.5}L)$,
where $T$ is the number of time slots, and $L$ is the maximum length
of the input. Although the combination of linear programming (LP)
and non-linear programming seems interesting, the algorithm is not
applicable to the general case of $\delta>1$ due to the prohibitively
high cost of the so-called pipage rounding for the general $\delta$
\cite{lu2014data}. Different to \cite{Infocom12LuEfficient} and
\cite{lu2014data}, the work \cite{He2013Efficient} converts the
relationship between data items and time slots to a directed acyclic
graph (DAG), and presents heuristic algorithms to compute a nearly
optimal access pattern for both one antenna and multiple antennae
scenarios. To the best of our knowledge, no approximation algorithm
is known for $\delta$ALWDR in the general case of $\delta>1$.

Since our algorithms have roots in the existing randomized algorithms
for the covering problem, the important results for the maximum coverage
problem (or namely, max $k-$cover), the minimum set cover (SC) problem,
etc. will be addressed. Essentially, LWDR can be considered as a maximum
coverage problem with restricts on the elements. The maximum coverage
problem is known to admit a ratio of $1-\frac{1}{e}$ that can be
achieved by a greedy algorithm \cite{hochbaum1996approximating}.
The key idea of the algorithm is always to select the set with maximum
uncovered weight, until all elements are covered. The ratio is best
possible, since this problem admits no ratio $1-\frac{1}{e}+\epsilon$
approximation even when all elements are with equal weight, under
the assumption that ${\cal P}\neq{\cal NP}$ \cite{feige1998threshold}.
It is interesting that, unlike the case for the maximum coverage problem,
applying similar idea as the greedy algorithm for $\delta$A$\gamma$LWDR
can only result in an approximation algorithm with a tight ratio $0.5$.
For a given collection $\mathbb{C}$ of subsets of $S=\{u_{1},\, u_{2},\,\dots,u_{n}\}$,
the minimum set cover (SC) problem is to compute a subset $\mathbb{C}'\subseteq\mathbb{C}$,
such that every element in $S$ belongs to at least one member of
$\mathbb{C}'$. It has been shown that SC can be approximated within
a factor of $\ensuremath{1+\ln\vert S\vert}$ \cite{johnson1973approximation}
and is not approximable within $c\log n$ unless ${\cal P}={\cal NP}$,
for some $c>0$ \cite{feige1998threshold}. When the cardinality of
all sets in $C$ are bounded by a given constant $k$, SC remains
${\cal APX}$-complete and is approximable within $\sum_{i=1}^{k}\frac{1}{i}-1/2$
\cite{duh1997approximation}. Moreover, if the number of occurrences
of any element in $\mathbb{C}$ is also bounded by a constant $c\geq2$,
SC remains ${\cal APX}$-complete \cite{papadimitriou1988optimization}
and approximable within a factor $c$ for both weighted and unweighted
SC \cite{bar1981linear,hochbaum1982approximation}.

In general, LWDR is to optimize a submodular function subject to a
number of constraints. For optimization of a submodular function subject
to candidate constraints, matroid constraints, or knapsack constraints,
the very recent results are approximation algorithms with ratio $1-\frac{1}{e}-\epsilon$,
for any fixed $\epsilon>0$ \cite{DBLP:conf/soda/BadanidiyuruV14}.
Their algorithms can not be applied to LWDR, since the constraints
therein are neither candidate constraints nor matroid constraints.

\subsection{Our Technique and Main Results}

In the paper, $\delta$ALWDR is first investigated and transferred
to the $\delta$-disjoint maximum weight longest path (with restricts)
problem in a layered DAG, and then a novel linear programming (LP)
formula for its relaxation is given accordingly. By the formula, we
show that $\delta$ALWDR is polynomial solvable when every data item
appears at most once. Contrastingly, general $\delta$ALWDR is ${\cal NP}$-hard
even when $\delta=1$. Thus, the difficulty of solving $\delta$ALWDR
mainly comes from the items with multiple occurrences. Further, the
LP formula is shown to be with an integrality gap 2. That means, immediately
based on the LP formula, it is impossible to develop approximation
with ratio better than $\frac{1}{2}$. Then a simple idea is first
to divide an instance of $\delta$ALWDR into a number of subinstances
in which every item appears at most once; then to solve the subinstances
individually, and to combine the computed subsolutions to a whole
solution. Following this idea, this paper develops three algorithms
approximating $\delta$A$\gamma$LWDR within a factor of $1-\frac{1}{e}$.
Based on the simple observation as in Proposition \ref{prop:simpobsbetw-gamma},
the algorithms can be extended for approximating $\delta$ALWDR within
a ratio $1-\frac{1}{e}-\epsilon$.

The first algorithm for $\delta$A$\gamma$LWDR is based on the randomized
LP rounding technique. The idea is inspired by the famous randomized
algorithms for set cover \cite{korte2002combinatorial}. Provided
that Karmarkar's algorithm is used to solve the LP formula \cite{korte2002combinatorial},
the algorithm is with a time complexity of $O(m^{3.5\gamma}\left(\frac{T}{\gamma}\right)^{3.5}L)$,
where $m$ is the number of channels, $\gamma$ is the factor of $\delta$A$\gamma$LWDR,
$T$ is the number of time slots, and $L$ is the maximum length of
the input. The algorithm can be extended to $\delta$ALWDR immediately,
within a runtime \textbf{$O(\epsilon^{3.5}m^{\frac{3.5}{\epsilon}}T^{3.5}L)$}
for any fixed $\epsilon>0$, which is the same as the result of \cite{lu2014data}
when $\delta=1$, but works for arbitrary $\delta$. We note that
the time complexity is high, e.g. it is $O(m^{35}T^{3.5}L)$ when
setting $\epsilon=0.1$.

Observe that the high runtime comes from the large number of all possible
paths involved in the first algorithm, we develop the second algorithm
via collectively randomized LP rounding technique for $\delta$A$\gamma$LWDR.
This is one of the main results of the paper. The algorithm is first
given and analyzed under the \textbf{occurrence assumption} (each
data item appears at most once in every segment of $\delta$A$\gamma$LWDR).
By collectively randomized LP rounding, our algorithm still randomly
rounds edges according to an optimum solution against the LP formula,
but simultaneously rounds up a collection of edges (a computed flow)
at one time instead of rounding the edges one by one individually.
The improved runtime of the algorithm is $O(m^{7}T^{3.5}L)$ using
Karmarkar's algorithm.\textbf{ }

Further, it is shown the second algorithm can be improved to solve
$\delta$A$\gamma$LWDR without the occurrence assumption, resulting
in the third algorithm of the same ratio $1-\frac{1}{e}$. The algorithm
can also be extended to approximate $\delta$ALWDR immediately, at
the cost of increasing runtime to $O(2^{\frac{1}{\epsilon}}\frac{1}{\epsilon}m^{7}T^{3.5}L)$.
This presents a significant improvement from the previous result $O(\epsilon^{3.5}m^{\frac{3.5}{\epsilon}}T^{3.5}L)$
for $\delta=1$ in \cite{lu2014data}. Although the improved time
complexity still looks high, we argue that it is efficient for two
reasons: (1) $m$ is not large in most cases (typically 2-20); (2)
for practical applications we may use the simplex method instead to
solve the LP formula. It is known that Karmarkar's algorithm (or other
interior-point method) has better worst case time complexity, but
the simplex method has a much better practical performance. As a by-product,
we also give a fixed-parameter tractable (fpt-)algorithm with a time
complexity $O(2^{B}m^{7}T^{3.5}L)$ for $\delta$ALWDR, where $B$
is the number of time slots that contain data items with multiple
occurrences.

In addition, we also prove the ${\cal NP}$-completeness of the restricted
version of $\delta$A$\gamma$LWDR with $\delta=1$ and $\gamma=2$,
by giving a reduction from the 3-dimensional perfect matching (3DM)
problem. We note that the ${\cal NP}$-completeness proof of LWDR
in \cite{Infocom12LuEfficient} can not be easily extended to show
the ${\cal NP}$-completeness of $\delta$A$\gamma$LWDR for $\gamma=2$
and $\delta=1$.

The remainder of this paper is organized as follows: Section II gives
first the construction of the DAG corresponding to $\delta$ALWDR,
then an LP-formula for the relaxation of the problem, as well as some
interesting properties; Section III presents a ratio $1-\frac{1}{e}$
randomized approximation algorithm with a time complexity $O(m^{3.5\gamma}\frac{T}{\gamma}^{3.5}L)$
for $\delta$A$\gamma$LWDR for $\delta=1$, as well as its ratio
proof and its extension to general $\delta$; Section IV gives a randomized
approximation algorithm with ratio $1-\frac{1}{e}$ and an improved
runtime $O(m^{7}T^{3.5}L)$ for $\delta$A$\gamma$LWDR under the
occurrence assumption, and then its derandomization; Section V shows
that the approximation algorithm can be improved such that it works
for $\delta$A$\gamma$LWDR with the same ratio $1-\frac{1}{e}$ but
without the occurrence assumption, and in addition that $\delta$ALWDR
is fixed-parameter tractable; Section VI gives the ${\cal NP}$-completeness
proof for $\delta$A$\gamma$-LWDR with $\gamma=2$ and $\delta=1$;
Section VII evaluates our algorithms by experiments; Section VIII
concludes this paper.

\section{DAG and LP Formula for $\delta$ALWDR}

This section will first transform an instance of $\delta$ALWDR into
a layered DAG $G$ with distinct vertices $s$ and $t$, such that
there exists a retrieve sequence for $\delta$ALWDR if and only if
there exist $\delta$ (edge) disjoint $st$-paths in $G$. Then an
LP formula is proposed for computing $\delta$ disjoint $st$-paths
in the constructed DAG, and hence for $\delta$ALWDR. Based on the
formula, $\delta$ALWDR is shown polynomial solvable when each data
item appears at most once in the time interval. Later, the LP formula
is shown with an integrality gap 2, so it is hard to approximate $\delta$ALWDR
within a factor better than 2 using the LP formula.

\subsection{Construction of the Auxiliary DAG }

The construction of DAG $G$ is as in Algorithm \ref{alg:1Construction-of-Auxiliary}.

\begin{algorithm}
\textbf{Input: }An instance of $\delta$ALWDR;

\textbf{Output:} $G$.
\begin{enumerate}
\item $G:=\emptyset$;
\item For every item $d_{i}$:

\begin{enumerate}
\item Add an edge set $E_{d_{i}}=\{e_{i,j,k}=(v_{i,\, j,\, k},\, w{}_{i,\, j,\, k})\vert d_{i}\mbox{ appears in the \ensuremath{j}th channel in the }k\mbox{th time slot}\}$
to $G$, where $w(e_{i,\, j,\, k})=w_{i}$;
\end{enumerate}
\item For the relationship between the items, add weight-0 edges to $G$
as below:

\begin{enumerate}
\item Edge $(w{}_{i,\, j,\, k},\, v_{i'\,,\, j',\, k+\Delta})$ to $G$
for every $i,\, i',\, j,\, j',\, k$ and every $\Delta\geq2$, $\Delta\in\mathbb{Z}$;
\item Edge $(w_{i,\, j,\, k},\, v_{i',\, j,\, k'})$ to $D$ for any $k'>k$;
/{*} For two items which are broadcast both in the $j$th channel.{*}/
\end{enumerate}
\item Add two vertices $s$ and $t$ with weight-0 edges to $G$ as below:

\begin{enumerate}
\item Edge $(s,\, v_{i,\, j,\, k})$ for every $i,$ $j$ and $k$;
\item Edge $(w{}_{i,\, j,\, k},\, t)$ for every $i$, $j$ and $k$.
\end{enumerate}
\end{enumerate}
\protect\caption{\label{alg:1Construction-of-Auxiliary}Construction of DAG for $\delta$ALWDR. }
\end{algorithm}

\begin{figure*}
\includegraphics{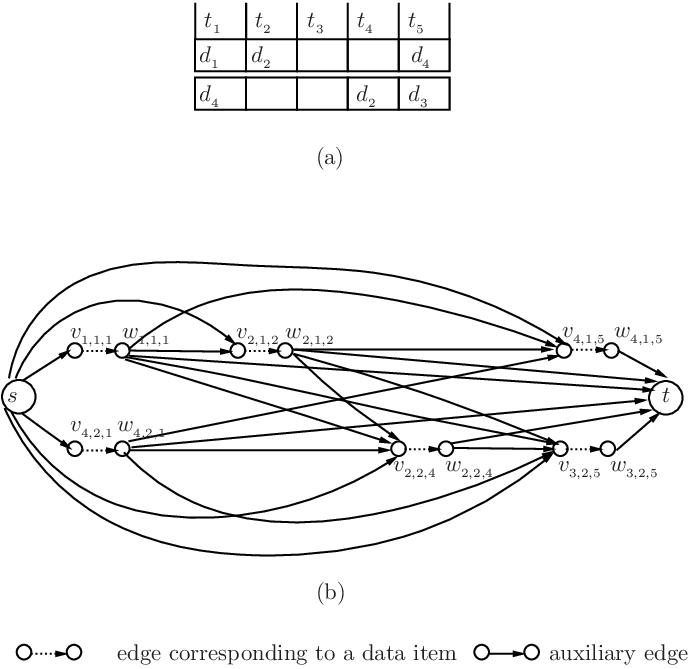}

\protect\caption{\label{fig:Construction-of-an}Construction of an auxiliary DAG for
an instance of LWDR: (a) An instance of LWDR; (b) The corresponding
auxiliary DAG. }
\end{figure*}

Note that $\vert E(G)\vert=O(\vert V(G)\vert^{2})$ according to the
above construction. Since $\vert V(G)\vert$ equals the number of
the occurrences of all the items, $\vert V(G)\vert=O(m\cdot T)$ holds,
where $m$ is the number of channels and $T$ is the number of time
slots. So $\vert E(G)\vert=O(m^{2}T^{2})$. However, it will be shown
later that $|E(G)|$ can be improved to $O(m\cdot\vert V(G)\vert)=O(m^{2}T)$.
Figure \ref{fig:Construction-of-an} depicts an example of constructing
a layered DAG for a given $\delta$ALWDR instance by Algorithm \ref{alg:1Construction-of-Auxiliary}.
For briefness, we say an instance of $\delta$ALWDR is feasible, if
and only if there exists a retrieve sequence according to which all
data items can be retrieved.
\begin{lem}
\label{lem:lwdtokpath}An instance of $\delta$ALWDR is feasible if
and only if there exist at most $\delta$ disjoint $st$-paths containing
at least one edge of each $E_{d_{i}}$ for each $i$ in the corresponding
DAG $G$.\end{lem}
\begin{IEEEproof}
We shall only show the lemma holds for the case $\delta=1$, since
the case for general $\delta$ is similar. Assume that $o_{i_{1},j_{1},k_{1}},\, o_{i_{2},j_{2},k_{2}},\dots,\, o_{i_{l},j_{l},k_{l}},\, o_{i_{l+1},j_{l+1},k_{l+1}},\,...$,
$k_{l+1}>k_{l}$ for any $l$, is a retrieve sequence for an instance
of LWDR, where $o_{i_{l},\, j_{l},\, k_{l}}$ is the occurrence of
data item $d_{i_{l}}$ in channel $j_{l}$ and time slot $k_{l}$.
If $o_{i_{l+1},\, j_{l+1},\, k_{l+1}}$ can be retrieved after $o_{i_{l},\, j_{l},\, k_{l}}$,
then the two data items must be conflict-free. That is, $o_{i_{l+1},\, j_{l+1},\, k_{l+1}}$
and $o_{i_{l},\, j_{l},\, k_{l}}$ are either (1) in the same channel,
i.e. $j_{l+1}=j_{l}$, and $k_{l+1}>k_{l}$; or (2) in different channels
and $k_{l+1}-k_{l}\geq2$. According to the construction of $G$,
there must be an edge leaving $w{}_{i_{l},\, j_{l},\, k_{l}}$, the
head of the edge corresponding to $o_{i_{l},\, j_{l},\, k_{l}}$,
and entering $v{}_{i_{l+1},\, j_{l+1},\, k_{l+1}}$, the tail of the
edge corresponding to item $o_{i_{l+1},\, j_{l+1},\, k_{l+1}}$. Therefore,
$P=s,\, v_{i_{1},j_{1},k_{1}},\, w_{i_{1},j_{1},k_{1}},\, v_{i_{2},j_{2},k_{2}},\, w_{i_{2},j_{2},k_{2}},\dots,\, t$
is an $st$ path in $G$.

Conversely, assume that there exists in $G$ a path $P=s,\, v_{i_{1},j_{1},k_{1}},\, w_{i_{1},j_{1},k_{1}},\, v_{i_{2},j_{2},k_{2}},\, w_{i_{2},j_{2},k_{2}},\dots,\, v_{i_{l},j_{l},k_{l}},$ $w_{i_{l},j_{l},k_{l}},\,
v_{i_{l+1},j_{l+1},k_{l+1}},\, w_{i_{l+1},j_{l+1},k_{l+1}},\,\dots,\, t$,
$k_{l+1}>k_{l}$, sharing at least one edge with each $E_{d_{i}}$.
According to the construction as Algorithm \ref{alg:1Construction-of-Auxiliary},
there exists an edge $(w{}_{i_{l},j_{l},k_{l}},\, v{}_{i_{l+1},j_{l+1},k_{l+1}})$
only if $k_{l+1}-k_{l}\geq2$ or $j_{l}=j_{l+1}$ and $k_{l+1}>k_{l}$.
That is, the items corresponding to $(v_{i_{l},j_{l},k_{l}},\, w_{i_{l},j_{l},k_{l}})$
and $(v_{i_{l+1},j_{l+1},k_{l+1}},\, w_{i_{l+1},j_{l+1},k_{l+1}})$,
say $o_{i_{l},\, j_{l},\, k_{l}}$ and $o_{i_{l+1},\, j_{l+1},\, k_{l+1}}$,
are conflict-free. That is, $o_{i_{1},j_{1},k_{1}},\, o_{i_{2},j_{2},k_{2}},\dots$
is a valid retrieve sequence. Then since $P$ contains at least one
edge of each $E_{d_{i}}$, the retrieve sequence retrieves all data
items.
\end{IEEEproof}
While no confusion arises, an edge is said in the $k$th time slot,
if the edge is corresponding to an item broadcast in the $k$th time
slot.

\subsection{An Linear Programming Relaxation for $\delta$ALWDR }

By Lemma \ref{lem:lwdtokpath}, to compute an optimum solution to
$\delta$ALWDR, we need only to compute $\delta$-disjoint maximum
weight $st$-paths (with some additional restricts) in $G$. The following
formula is an LP relaxation for $\delta$ALWDR (an LP relaxation for
$\delta$ALNDR if $w(e)=1$ for all $e\in E$):

\begin{eqnarray}
max &  & \sum_{e\in E}w(e)x_{e}\label{eq:theoriginalLP}\\
s.t. &  & \sum_{e\in E_{d_{i}}}x_{e}\leq1\begin{array}{cc}
 & \forall E_{d_{i}}\subseteq E\end{array}\label{eq:keyconstr}\\
 &  & \sum_{e\in\delta^{+}(v)}x_{e}-\sum_{e\in\delta^{-}(v)}x_{e}=\left\{ \begin{array}{cc}
0 & \mbox{ }\forall v\in V\setminus\{s,\, t\}\\
\delta & v=s
\end{array}\right.\label{eq:unimo}\\
 &  & 0\leq x_{e}\leq1\mbox{ }\forall e\in E(G)\label{eq:constrxe}
\end{eqnarray}

If Inequality (\ref{eq:keyconstr}) is removed, then the above formula
is exactly an LP formula for the relaxation of $\delta$-disjoint
$st$-paths in $G$. Because graph $G$ is acyclic, any integral optimum
solution (i.e. a solution with all $x_{e}\in\{0,1\}$) to LP (\ref{eq:theoriginalLP})
contains no cycle. Thus, the solution is a set of (edge) disjoint
paths with maximum weight, and hence an optimum solution to $\delta$ALWDR.
Note that the condition that graph $G$ is acyclic is essential. Otherwise,
a solution to LP (\ref{eq:theoriginalLP}) can contain both cycles
and paths, and is not a solution to $\delta$ALWDR.

Further, since when $\vert E_{d_{i}}\vert=1$ holds for each $d_{i}$,
the constraint matrix of LP (\ref{eq:theoriginalLP}) is known totally
unimodular \cite{schrijver1998theory}, we have the following property
that indicates $\delta$ALWDR is polynomial solvable when each data
item broadcast at most once:
\begin{thm}
\label{Thr:LWDR-is-polynomial-1} When $\vert E_{d_{i}}\vert=1$ holds
for each $d_{i}$, any basic optimum solution to LP (\ref{eq:theoriginalLP})
is integral, i.e., each edge $e$ is with $x_{e}=0$ or $x_{e}=1$.
\end{thm}

\begin{figure}
\includegraphics{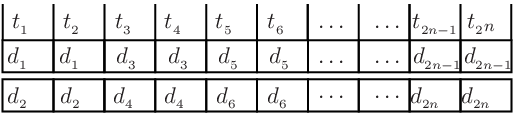}

\protect\caption{\label{fig:Integrality-Gap-of}Integrality gap of LP (\ref{eq:theoriginalLP}). }
\end{figure}

Therefore, according to the theorem above, to solve an instance of
$\delta$ALWDR in which each item appears at most once, we need only
to compute a basic optimum solution to LP (\ref{eq:theoriginalLP})
with $\forall i$ $\vert E_{d_{i}}\vert=1$. Moreover, it is known
that a basic optimum solution to LP (\ref{eq:theoriginalLP}) can
be computed in polynomial time \cite{korte2002combinatorial}. Hence,
we have:
\begin{cor}
\label{cor:LWDRpolynomial}$\delta$ALWDR is polynomial solvable when
each data item is broadcast at most once.
\end{cor}
However, it is known the general case of $\delta$ALWDR is ${\cal NP}$-hard
even when $\delta=1$. Worse still, it is hard to approximate $\delta$ALWDR
within a factor better than $\frac{1}{2}$ via LP (\ref{eq:theoriginalLP}),
as stated in the following observation:
\begin{prop}
\label{prop:The-integrality-gap}The integrality gap of LP(\ref{eq:theoriginalLP})
is $\frac{1}{2}$, even for LNDR with only two channels and each data
item is broadcast at most twice.
\end{prop}
To show the integrality gap as above, we give an instance of LWDR
as depicted in Figure \ref{fig:Integrality-Gap-of}, where there are
only two channels, one antenna, and every item in the instance has
the same weight and is broadcast at most twice. Then apparently, an
optimal solution is to retrieve data items $\{d_{i}\vert i=1,\,3,\,\dots,\,2n-1\}$, for
which the antenna only needs to keep listening to the upper channel.
That is, the weight of the retrieved data items of an optimal solution
is $n$. On the other hand, $\{x_{e}=\frac{1}{2}\vert e\in E_{d_{i}}\}$
is an optimal solution to LP (\ref{eq:theoriginalLP}) against the
instance, resulting a weight of 2$n$.

Then following the definition of integrality gap, it is impossible
to design an approximation algorithm with ratio better than 2 based
on LP (\ref{eq:theoriginalLP}) (using LP-rounding, primal-dual method,
etc). However, it is worth noting that the integrality gap of LP (\ref{eq:theoriginalLP})
is better than 2 if the occurrence assumption holds. In fact, that
is why we propose the occurrence assumption.

\section{A Factor$-1-\frac{1}{e}$ Approximation Algorithm for $\delta$A$\gamma$LWDR}

Instead of approximating $\delta$ALWDR directly, this section will
first give an approximation algorithm with ratio $1-\frac{1}{e}$
for $\delta$A$\gamma$LWDR for $\delta=1$, and then show that it
can be extended to general $\delta$. The key idea of the algorithm
comes from the following simple observation that can be easily extended
from a lemma in\cite{lu2014data}.
\begin{prop}
All possible $k$ retrieval sequences can be computed in $O(m^{kT})$
time for $\delta$ALWDR, where $T$ is the number of the time slots.
\end{prop}
From the above proposition, a simple idea is first to solve each segment
of $\delta$A$\gamma$LWDR individually (since a segment is an instance
of $\delta$ALWDR with $\gamma$ time slots), and then to combine
the computed subsolutions to a whole solution. However, the difficulty
is how to guarantee that the subsolutions could compose a good solution.
To overcome the difficulty, we first compute all possible paths for
each segment of $\delta$A$\gamma$LWDR, then give a LP formula for
$\delta$A$\gamma$LWDR based on the set of paths. Based on the formula,
an approximation algorithm is developed by employing randomized rounding
technique. The approximation achieves the same ratio of $1-\frac{1}{e}$ as
in \cite{lu2014data}, but is simpler and can be easily extended to
solve $\delta$A$\gamma$LWDR (and $\delta$ALWDR) for general $\delta$.

\subsection{The LP Formula for Relaxation of $\delta$A$\gamma$LWDR}

Let $G$ be the auxiliary graph output by Algorithm \ref{alg:1Construction-of-Auxiliary}.
Let $t_{j_{0}}=t_{0},t_{j_{1}},\dots,t_{j{}_{l}}\dots$ be the vacant
time slots in $\delta$A$\gamma$LWDR. Assume that $G_{i}$ is the
part of $G$ between $t_{j_{i-1}}$ and $t_{j_{i}}$, i.e. it corresponds
to the $i$th segment of $\delta$A$\gamma$LWDR. Let $\mathbb{P}=\{\mathbb{P}_{1},\dots,\mathbb{P}_{N}\}$,
where $\mathbb{P}_{i}=\{P_{i,j}\vert j=1,\,\dots,\, h_{i}\,\mbox{and}\,\vert P_{i,j}\cap E_{d_{l}}\vert\leq1\,\forall l\}$
is the set of all the possible paths of $G_{i}$ that correspond to
retrieve sequences. Each $P_{i,j}$ is assigned with a weight $w(P_{i,j})=\sum_{e\in P_{i,j}}w(e)$.
Formally, our LP relaxation for $\delta$A$\gamma$LWDR is as below:

\begin{eqnarray}
max &  & \sum_{i=1}^{N}\sum_{P_{i,j}\in\mathbb{P}_{i}}w(P_{i,j})\cdot x_{i,j}\label{eq:LP}\\
s.t. &  & \sum_{P_{i,j}\in\mathbb{P}_{i}}x_{i,j}=\delta\quad\forall\mathbb{P}_{i}\in\mathbb{P}\label{eq:unimoforSC}\\
 &  & \sum_{i,j:\, P_{i,\, j}\cap E_{d_{l}}\neq\emptyset}x_{i,j}\leq1\quad\forall E_{d_{l}}\subseteq E(G)\label{eq:edl}\\
 &  & \quad\quad0\leq x_{i,\, j}\leq1\quad\forall P_{i,j}\in\mathbb{P}_{i}
\end{eqnarray}

In the above formula, $x_{i,j}=1$ indicates $P_{i,j}$ is selected,
and $x_{i,j}=0$ otherwise. Inequality (\ref{eq:edl}) is to guarantee
that a feasible solution of LP (\ref{eq:LP}) contains at most one
edge of each $E_{d_{l}}$, i.e. at most one edge for item $d_{l}$.
If $x_{i,j}\in\{0,1\}$ $1$, the above formula becomes an integral
programming (IP) formula for $\delta$A$\gamma$LWDR.

\subsection{The Randomized Algorithm for $\delta$A$\gamma$LWDR }

Let $\chi=(x_{1,1}^{*},\,\dots,x_{i,j}^{*},\dots)$ be an optimal
solution to LP (\ref{eq:LP}), and $w_{LP}$ be its weight. The key
idea of our algorithm is to interpret the fractional value $x_{i,j}^{*}$
as the probability of selecting $P_{i,j}$ for $\mathbb{P}_{i}$.
Then the algorithm is formally as in Algorithm \ref{alg:SCbased-randomized-algorithm}.

\begin{algorithm}
\textbf{Input: }$G$, $\mathbb{P}=\{\mathbb{P}_{1},\dots,\mathbb{P}_{N}\}$,
where $\mathbb{P}_{i}=\{P_{i,j}\vert j=1,\,\dots,\, h_{i}\}$ is a
collection of paths of $G_{i}$, with a weight $w(P_{i,j})=\sum_{e\in P_{i,j}}w(e)$;

\textbf{Output:} $\mathbb{C}$, a solution to $\delta$A$\gamma$LWDR
for $\delta=1$.
\begin{enumerate}
\item $\mathbb{C}:=\emptyset$;
\item Solve LP (\ref{eq:LP}) against $\mathbb{P}$ for $\delta=1$ by Karmarkar's
algorithm \cite{schrijver1998theory}, and obtain an optimal solution
$\mathbf{x}=(x_{1,1}^{*},\,\dots,x_{i,j}^{*},\dots)$;
\item \textbf{For} $i=1$ to $N$ \textbf{do }

\begin{enumerate}
\item Set $P_{i}:=P_{i,\, j}$ with probability $x_{i,j}^{*}$; /{*} $P_{i}$
is the element selected in $\mathbb{P}_{i}$. {*}/
\item $\mathbb{C}:=\mathbb{C}\cup\{P_{i}\}$;
\end{enumerate}
\item Return $\mathbb{C}$.
\end{enumerate}
\protect\caption{\label{alg:SCbased-randomized-algorithm}A randomized algorithm for
$\delta$A$\gamma$LWDR.}
\end{algorithm}

\begin{lem}
\label{lem:SCratioproof}Algorithm \ref{alg:SCbased-randomized-algorithm}
is a randomized $(1-(\frac{K-1}{K})^{K})$-approximation algorithm
with a time complexity of $O(m^{3.5\gamma}\cdot(\frac{T}{\gamma})^{3.5}\cdot L)$
for $\delta$A$\gamma$LWDR for $\delta=1$, where $L$ is the maximum
length of input and $K$ is the maximum occurrence times of $e$ in
all $P_{i,j}$. \end{lem}
\begin{IEEEproof}
The runtime of Algorithm \ref{alg:SCbased-randomized-algorithm}
is easy to calculate: Step 2 of the algorithm takes $O\left(\left(\sum_{i=1}^{N}\vert\mathbb{P}{}_{i}\vert\right)^{3.5}L\right)$
time to run Karmarkar's algorithm, where $\vert\mathbb{P}_{i}\vert=O\left(m^{\gamma}\right)$.
Then because other steps take trivial time compared to Step 2, the
total time is $O(\left(N\cdot m^{\gamma}\right)^{3.5}L)=O(m^{3.5\gamma}\cdot(\frac{T}{\gamma})^{3.5}\cdot L)$.

For the ratio, let $SOL$ and $w_{SOL}$ be the output of the algorithm
and its weight respectively. To calculate the expected value of the
output of the algorithm, it remains only to compute the probability
that none of the edges of $E_{d_{l}}$ is in any $P_{i,j}\in SOL$.
Below is the probability that $E_{d_{l}}\cap P_{i,j}=\emptyset$ for
every $P_{i,j}\in SOL$:

\[
\prod_{i}\left(1-\sum_{j:\, E_{d_{l}}\cap P_{i,,j}\neq\emptyset}x_{i,j}^{*}\right)\leq\prod_{i,j:\, E_{d_{l}}\cap P_{i,,j}\neq\emptyset}\left(1-x_{i,j}^{*}\right).
\]

Then since $\sum_{i,j:\, E_{d_{l}}\cap P_{i,,j}\neq\emptyset}x_{i,j}^{*}$
is fixed, we assume that \[f=\prod_{i,j:\, E_{d_{l}}\cap P_{i,,j}\neq\emptyset}(1-x_{i,j}^{*})-\lambda\cdot\sum_{i,j:\, E_{d_{l}}\cap P_{i,,j}\neq\emptyset}x_{i,j}^{*}\].
It is easy to see $f$ attains maximum when

\[
\frac{\partial f}{\partial x_{i,j}^{*}}=\lambda\mbox{ }\forall i,j:\, u_{k}\in P_{i,j}.
\]

That is, when all elements of $\left\{ x_{i,j}^{*}\vert E_{d_{l}}\cap P_{i,,j}\neq\emptyset\right\} $
are the same, $\prod_{i,j:\, E_{d_{l}}\cap P_{i,,j}\neq\emptyset}(1-x_{i,j}^{*})$
attains maximum $(1-\frac{1}{K})^{K}$, where $K$ is the number of
the occurrence times of any edge of $E_{d_{l}}$ appearing in all
$P_{i,j}$. So edges of $E_{d_{l}}$ have a probability of at most
$(1-\frac{1}{K})^{K}$ to be all absent from every $P_{i,j}$. Now
we have all the ingredients for computing $E(w_{SOL})$, the expectation
of $w_{SOL}$:
\begin{eqnarray}
E(w_{SOL}) & = & w_{LP}-(\mbox{expected weight of the elements absenting every }P_{i,j})\cdot w_{LP}\label{eq:optlp}\\
 & \geq & (1-(1-\frac{1}{K})^{K})\cdot w_{LP}\nonumber
\end{eqnarray}
Since $w_{LP}$ is the weight of an optimal solution to LP (\ref{eq:LP}),
$w_{LP}$ is not less than $w_{OPT}$, the weight of an optimal solution
to $\delta$A$\gamma$LWDR. Therefore, $E(w_{SOL})\geq(1-(1-\frac{1}{K})^{K})\cdot w_{OPT}$.
This completes the proof.
\end{IEEEproof}
By simple arithmetical calculation, it is easy to see the ratio will
be $0.75$ when $K=2$, be $0.704$ when $K=3$, and be $0.684$ when
$K=4$. Further, following the inequality as in Proposition \ref{prop:calK}
below, the ratio of our algorithm would be not less than $\lim_{K\rightarrow+\infty}1-(\frac{K-1}{K})^{K}=1-\frac{1}{e}$.
\begin{prop}
\label{prop:calK}$f(K)=1-(\frac{K-1}{K})^{K}$ is a monotone increasing
function for $K\geq2$.\end{prop}
\begin{IEEEproof}
The derivative of $f(K)$ is as in the following:

\[
f(K)'=((\frac{K-1}{K})^{K})'=(e^{K(\ln(K-1)-\ln K)})'=e^{K(\ln(K-1)-\ln K)}\cdotp(\frac{1}{k-1}+\ln\frac{K-1}{K}).
\]
Because for $K=2$, we have both $\frac{1}{k-1}+\ln\frac{K-1}{K}>0$
and its derivative $(\frac{1}{k-1}+\ln\frac{K-1}{K})'=\frac{1}{(K-1)^{2}}>0$,
$f(K)'>0$ holds for any $K\geq2$. That is, for any $K\geq2$, $f(K)$
is monotone increasing.
\end{IEEEproof}
The derandomization of Algorithm \ref{alg:SCbased-randomized-algorithm}
to $\delta$A$\gamma$LWDR follows a similar line as the derandomization
of Section 4 (although it is a little more complicated). So we omit
it here.

\subsection{Extension to $\delta$A$\gamma$LWDR}

In this subsection, Algorithm \ref{alg:SCbased-randomized-algorithm}
is extended to solve $\delta$ALWDR for general $\delta$. To do this,
two changes are needed for the algorithm: the first is to change the
formula of LP (\ref{eq:LP}), by setting the right part of the constraint
of Equality (\ref{eq:unimoforSC}) from 1 to $\delta$ accordingly;
the second is to select $\delta$ disjoint paths to round up simultaneously
for each $\mathbb{P}_{h}$, while Algorithm \ref{alg:SCbased-randomized-algorithm}
round up only one path for each $\mathbb{P}_{h}$. More precisely,
the second is to modify Step 3 of Algorithm \ref{alg:SCbased-randomized-algorithm}
as below:

\textbf{For} each $\{P_{h,j_{1}},\dots,P_{h,j_{\delta}}\}\subseteq\mathbb{P}_{h}$
with $P_{h,j_{i}}\cap P_{h,j_{k}}=\emptyset$ for \textbf{$j_{i}\neq j_{k}$
do }

\quad{}(a) Set $\mathbb{{\cal P}}_{h}:=\{P_{h,j_{1}},\dots,P_{h,j_{\delta}}\}$
with probability $\frac{\sum_{l=1}^{\delta}x_{i,j_{l}}^{*}}{\delta}$,
where $\{P_{h,j_{1}},\dots,P_{h,j_{\delta}}\}$ is a set of $\delta$
disjoint paths in $\mathbb{P}_{h}$ and $\mathbb{{\cal P}}_{h}$ is
the set of $\delta$ paths selected from $\mathbb{P}_{h}$.

\quad{}(b) $\mathbb{C}:=\mathbb{C}\cup\mathbb{{\cal P}}_{h}$.
\begin{lem}
\label{lem:aArLWDRratioandtime}$\delta$A$\gamma$LWDR admits an
approximation algorithm with ratio $1-\frac{1}{e}$ and runtime $O(m^{3.5\gamma}\cdot(\frac{T}{\gamma})^{3.5}\cdot L+\frac{T}{\gamma}m^{\delta\gamma})$.\end{lem}
\begin{IEEEproof}
The ratio can be obtained following exactly the same line of Lemma
\ref{lem:SCratioproof}.

For the time complexity, Step 2 takes $O(m^{3.5\gamma}\cdot(\frac{T}{\gamma})^{3.5}\cdot L)$
time to solve LP (\ref{eq:LP}). Step 3 takes $\left(\begin{array}{c}
m^{\gamma}\\
\delta
\end{array}\right)=O(m^{\delta\gamma})$ time to select a ${\cal P}_{h}$ to round up, since it has to select
$\delta$ paths among $m^{\gamma}$ paths which are all possible paths
in $\mathbb{P}_{i}$. Therefore, the total runtime is $O(m^{3.5\gamma}\cdot(\frac{T}{\gamma})^{3.5}\cdot L+\frac{T}{\gamma}m^{\delta\gamma})$.
This completes the proof.
\end{IEEEproof}

\section{Approximation Algorithms for $\delta$A$\gamma$LWDR under Occurrence
Assumption}

The section will give an algorithm to approximate $\delta$A$\gamma$LWDR
within a factor of $1-\frac{1}{e}$ for general $\delta$, under the
\textbf{occurrence assumption} that every item is broadcast at most
once in each segment. To do this, an approximation algorithm is first
given for $\delta$A$\gamma$LWDR for $\delta=1$, with the key idea
of collectively and randomly rounding fractional edges according to
an optimum solution to LP (\ref{eq:theoriginalLP}). Later, the algorithm
is derandomized using an interesting method based on conditional expectation.

For all the algorithms in this section, the key observation is that
the high time complexity of Algorithm \ref{alg:SCbased-randomized-algorithm}
mainly comes from the large size of $\mathbb{P}_{i}$, $m^{\gamma}$,
where $m$ is the number of the channels and $\gamma$ the length
of a segment. The basic idea of the algorithms is to compute the necessary
paths only, instead of computing all possible paths. To do this, we
release paths (or more precisely subflows) from the fractional flow
of an optimum solution to LP (\ref{eq:theoriginalLP}). Those released
paths compose the set of necessary paths for each $\mathbb{P}_{i}$.
An integral solution to $\delta$A$\gamma$LWDR can then be obtained
by rounding such fractional paths.

\subsection{A Randomized Algorithm for $\delta$A$\gamma$LWDR under Occurrence
Assumption}

Our algorithm is mainly composed by the following steps: first to
compute an optimum solution against LP (\ref{eq:theoriginalLP}) for
the constructed graph $G$, output by Algorithm \ref{alg:1Construction-of-Auxiliary}
for an given $\delta$A$\gamma$LWDR instance; then for each $\gamma$-separated
segment, the algorithm releases a set of subflows with fractional
value from the computed solution; later, the value of each subflow
is randomly rounded to 1 with probability proportional to the original
value. Eventually, the combination of the rounded subflows of each
$\gamma$-separated segment will collectively compose an integral
solution to $\delta$A$\gamma$LWDR. The full layout of the algorithm
is as in Algorithm \ref{alg:theflows-randomized-algorithm}.

\begin{algorithm}
\textbf{Input: }Auxiliary graph $G$ corresponding to an instance
of $\delta$A$\gamma$LWDR with occurrence assumption and $\delta=1$,
in which $G_{i}$ is corresponding for the $i$th segment of $\delta$A$\gamma$LWDR;

\textbf{Output:} A solution to $\delta$A$\gamma$LWDR.
\begin{enumerate}
\item Solve LP (\ref{eq:theoriginalLP}) against $G$ by Karmarkar's algorithm
\cite{schrijver1998theory}, and obtain an optimal solution $\mathbf{x}=(x_{1,1,1}^{*},\,\dots,x_{i,j,k}^{*},\dots)$;
\item \textbf{For} $h=1$ to $N$ \textbf{do}

\begin{enumerate}
\item For the flow corresponding to $\mathbf{x}$, divide the part in $G_{h}$
into a set of subflows, say $\mathbb{F}_{h}=\{f_{h,1},\,\dots,\, f_{h,\, j_{h}}\}$
where $f_{h,\, j}$ is with value $y_{h,\, j}$, $j\in\{1,\,\dots,\, j_{h}\}$ by
using Algorithm \ref{alg:Comp_Fh} (given later);
\item Round the value of $f{}_{h,\, j}$ to 1 with probability $y_{h,\, j}$;
\end{enumerate}
\item Return the set of the subflows with value 1 as a solution to $\delta$A$\gamma$LWDR.
\end{enumerate}
\protect\caption{\label{alg:theflows-randomized-algorithm}A collective randomized
rounding algorithm for $\delta$A$\gamma$LWDR.}
\end{algorithm}

\begin{lem}
\label{lem:flowOAtime}Algorithm \ref{alg:theflows-randomized-algorithm}
outputs a solution for $\delta$A$\gamma$LWDR within runtime $O(\vert E(G)\vert{}^{3.5}L)$,
where $L$ is the maximum length of the input.\end{lem}
\begin{IEEEproof}
Step 1 of the algorithm runs Karmarkar's algorithm and takes $O(\vert E(G)\vert{}^{3.5}L)$
time to solve LP(\ref{eq:LP}) wrt the auxiliary graph $G$, since
the number of the constraints of LP (\ref{eq:LP}) is $O(\vert E(G)\vert+\vert V(G)\vert)=O(\vert E(G)\vert)$.
According to Lemma \ref{lem:ratioOAflowLPadmits} (which is given
later), Step 2 takes $O(\vert E(G_{h})\vert^{2})$ time to compute
each $\mathbb{F}_{h}$, and the rounding time for each $\mathbb{F}_{h}$
is $O(\vert\mathbb{F}_{h}\vert)=O(\vert E(G_{h})\vert^{2})$. So the
total time of Step 2 is $O(N\cdot\vert E(G_{h})\vert^{2})$. Therefore,
the total time complexity of Algorithm \ref{alg:theflows-randomized-algorithm}
is $O(\vert E(G)\vert{}^{3.5}L)$.
\end{IEEEproof}
Let $m$ be the number of channels and $T$ be the number of time
slots. Recall that $\vert V(G)\vert$ equals to the number of all
the occurrences of the data items $O(mT)$, and $\vert E(G)\vert=O(\vert V(G)\vert^{2})$
according to the construction of $G$ as Algorithm \ref{alg:1Construction-of-Auxiliary}.
However, as will be shown in Section V, $\vert E(G)\vert$ can be
decreased to $O(m\cdot\vert V(G)\vert)=O(m^{2}T)$. So the runtime
of Algorithm \ref{alg:theflows-randomized-algorithm} is actually
$O(m^{7}T^{3.5}L)$.
\begin{lem}
\label{lem:ratioOAflowLPadmits}Algorithm \ref{alg:theflows-randomized-algorithm}
is a randomized $(1-(\frac{K-1}{K})^{K})$-approximation algorithm
for $\delta$A$\gamma$LWDR, where $K$ is the maximum occurrence
times of $d_{i}$ in the subflows in the time interval. \end{lem}
\begin{IEEEproof}
Let $SOL$ and $OPT$ be the output of the algorithm and the optimum
solution of $\delta$A$\gamma$LWDR, respectively. We shall show $E(w(SOL))\geq w(OPT)\cdot(1-(\frac{K-1}{K})^{K})$,
where $E(w(SOL))$ is the expectation of $w(SOL)$. Let $w_{LP}$
be the weight of an optimum solution to LP (\ref{eq:theoriginalLP})
against $G$. Since $w(OPT)\leq w_{LP}$, it remains only to show
$E(w(SOL))\geq w_{LP}\cdot(1-(\frac{K-1}{K})^{K})$. Thus, we will
calculate how much weight is expected to lose during the rounding
procession. For item $d_{i}$, assume that
\begin{eqnarray*}
\sum_{j,k,h,l:\, e_{i,j,k}\in f_{h,l}}y_{h,l} & = & \alpha,\,\,\,\,\,\,0\leq\alpha\leq1,
\end{eqnarray*}
where $y_{h,l}$ is the value of flow $f_{h,l}$ which contains $e_{i,j,k}$.
Then the probability that the algorithm does not pick $d_{i}$ is:

\[
(1-\alpha)+\alpha\cdot\prod_{j,k,h,l:\, e_{i,j,k}\in f_{h,l}}(\alpha-y_{h,l}).
\]

Then since $\sum_{j,k,h,l:\, e_{i,j,k}\in f_{h,l}}y_{h,l}=\alpha$
is fixed, we assume that $g_{i}=\prod_{j,k,h,l:\, e_{i,j,k}\in f_{h,l}}(\alpha-y_{h,l})-\lambda\cdot\sum_{j,k,h,l:\, e_{i,j,k}\in f_{h,l}}y_{h,l}$.
Similar to the proof of Lemma \ref{lem:SCratioproof}, it is easy
to see $g_{i}$ attains maximum when

\[
\frac{\partial g_{i}}{\partial y_{h,l}}=0\mbox{ }\forall j,k,h,l:\, e_{i,j,k}\in f_{h,l}.
\]

That is, $g_{i}$ attains maximum when every $y_{h,l}:\, f_{h,l}\supseteq\{e_{i,j,k}\}$
is with the same value. Then the probability that the algorithm does
not pick any edge corresponding to $d_{i}$ is at most:

\[
(1-\alpha)+\alpha\cdot(1-\frac{1}{K})^{K},
\]

where $K$ is the number of the occurrences of $e_{i,j,k}$, $\forall j,\, k$,
in all subflows, i.e. $f_{h,\, l}$, $\forall h,\, l$. So the expectation
of item $d_{i}$ being picked is $\alpha(1-(1-\frac{1}{K})^{K})$.
Therefore the expectation of $w(SOL)$ is:
\begin{eqnarray*}
E(w(SOL)) & = & w_{LP}\left(\frac{\mbox{the probability of the edges responding to \ensuremath{d_{i}}appearing in any }f_{l}}{\alpha}\right)\\
 & \geq & (1-(1-\frac{1}{K})^{K})\cdot w_{LP}
\end{eqnarray*}
This completes the proof.
\end{IEEEproof}
It remains to give the division of the subflows for Algorithm \ref{alg:theflows-randomized-algorithm}.
Let $\mathbf{x}=(x_{1,1,1}^{*},\,\dots,x_{i,j,k}^{*},\,\dots)$ be
an optimal solution to LP (\ref{eq:LP}). W.l.o.g., assume that we
are processing the edges of $G_{h}$, and $E_{h}=\{e\vert0<x_{e}\leq1\}$,
i.e. $E_{h}$ is the set of edges with $0<x_{e}\leq1$ in $G_{h}$.
The key idea of the computation is to repeatedly select an edge $e$
with minimum $x_{e}$, and then construct in $E_{h}$ a flow (which
is also a single path) of value $x_{e}$ going through $e$, until
every edge $e$ in $E_{h}$ is with $x_{e}=0$. The detailed algorithm
is shown in Algorithm \ref{alg:Comp_Fh}.

\begin{algorithm}
\textbf{Input}: $G_{h}$, $E_{h}$ and $\mathbf{x}=(x_{1,1,1}^{*},\,\dots,x_{i,j,k}^{*},\dots)$,
an optimum solution to LP (\ref{eq:theoriginalLP});

\textbf{Output}: $\mathbb{F}_{h}$.
\begin{enumerate}
\item Set $\mathbb{F}_{h}:=\emptyset$, $i:=1$;
\item Set $e^{*}:=nil$, $x{}_{e^{*}}:=0$, $f_{h,i}:=\emptyset$;
\item \textbf{For }each $e\in E_{h}$ \textbf{do}

\begin{enumerate}
\item \textbf{If} $x_{e}<x_{e^{*}}$ \textbf{then}

$e^{*}:=e$;

\end{enumerate}

/{*}Find an edge $e^{*}\in E_{h}$ with $x_{e^{*}}\leq x_{e}$ for
any $e\in E_{h}$. {*}/

\item $f_{h,i}:=f_{h,i}\cup\{e^{*}\}$;
\item Set $e_{pre}:=e_{suc}:=e^{*}$;
\item \textbf{While} $e_{pre}$ has preceding edges \textbf{do}

\begin{enumerate}
\item Select one of the preceding edges, say $e$;
\item Set $x_{e}:=x_{e}-x_{e^{*}}$, $e_{pre}:=e$, and $f_{h,i}:=f_{h,i}\cup\{e^{*}\}$;
\end{enumerate}

\textbf{EndWhile}

/{*}Add the part of $f_{h,i}$ before $e^{*}$ to $f_{h,i}$.{*}/

\item \textbf{While} $e_{suc}$ has successor edges \textbf{do}

\begin{enumerate}
\item Select one of the successor edges, say $e$;
\item Set $x_{e}:=x_{e}-x_{e^{*}}$, $e_{suc}:=e$, and $f_{h,i}:=f_{h,i}\cup\{e_{suc}\}$;
\end{enumerate}

\textbf{EndWhile}

/{*}Add the part of $f_{h,i}$ after $e^{*}$ to $f_{h,i}$.{*}/

\item Set $x_{e^{*}}:=0$ and $\mathbb{F}_{h}:=\mathbb{F}_{h}\cup f_{h,i}$;
/{*}Add $f_{h,i}$ to $\mathbb{F}_{h}$.{*}/
\item \textbf{For} each $e\in E_{h}$ \textbf{do}

$\quad$If $x_{e}=0$ \textbf{then} $E_{h}:=E_{h}\setminus\{e\}$;

\item \textbf{If} $E_{h}\neq\emptyset$ \textbf{then}

$\quad$Set $i:=i+1$ and go to Step 2;

\textbf{Else} return $\mathbb{F}_{h}$.

\end{enumerate}
\protect\caption{\label{alg:Comp_Fh}Computation of $\mathbb{F}_{h}$.}
\end{algorithm}

\begin{lem}
\label{lem:timeofcompFh}Algorithm \ref{alg:Comp_Fh} runs in $O(\vert E(G_{h})\vert^{2})$
time, and correctly computes a set of flows $\mathbb{F}_{h}=\{f_{h,i}\}$
for $G_{h}$, such that $\vert\mathbb{F}_{h}\vert\leq\vert E(G_{h})\vert$
and $x_{e}=\sum_{h,i:\, e\in f_{h,i}}y_{h,i}$ holds for each $e\in G_{h}$.\end{lem}
\begin{IEEEproof}
Algorithm \ref{alg:Comp_Fh} iterates Step 2-10 at most $\vert E_{h}\vert$
times, since each iteration removes at least one edge from $E_{h}$.
In each iteration, Step 6 and 7 need to verify all edges of $E_{h}$
in the worst case, which is at most $O(E_{h})$. So Algorithm \ref{alg:Comp_Fh}
runs in $O(\vert E(G_{h})\vert^{2})$ time.

For $\vert\mathbb{F}_{h}\vert$, according to Algorithm \ref{alg:Comp_Fh},
the size of $E_{h}$ decreases at least one when a new flow is added
to $\mathbb{F}_{h}$, since at least an edge $e^{*}$ with its $x_{e^{*}}$
set to 0 is removed from $E_{h}$. So $\vert\mathbb{F}_{h}\vert\leq\vert E_{h}\vert\leq\vert E(G_{h})\vert$.
For the latter part, $x_{e}=\sum_{h,i:\, e\in f_{h,i}}y_{h,i}$ clearly
holds since Algorithm \ref{alg:Comp_Fh} decreases $y_{h,i}$ from
$x_{e}$ if and only if $e\in f_{h,i}$.
\end{IEEEproof}
Similar to the extension of Algorithm \ref{alg:SCbased-randomized-algorithm}
to $\delta$A$\gamma$LWDR, and the analysis of Lemma \ref{lem:aArLWDRratioandtime},
it is easy to extend Algorithm \ref{alg:theflows-randomized-algorithm}
to $\delta$A$\gamma$LWDR for general $\delta$. Hence, the extension
is omitted.

\subsection{Derandomization}

The main idea of our derandomization is inspired by the derandomization
technique using conditional expectations as implicitly given in \cite{erdos1973combinatorial}
and formally given in the book \cite{spencer1987ten}. That is, to
pick $f_{i,j}$ for $\mathbb{F}_{i}$ in a greedy and sequential way:
the flow for $\mathbb{F}_{1}$ is first to select, then $\mathbb{F}_{2}$,
$\mathbb{F}_{3}$, $\dots$, and so on. Assume that the selection
of the flow for $\mathbb{F}_{1},\dots,\mathbb{F}_{h}$ is complete,
and the algorithm is currently selecting flow for $\mathbb{F}_{h+1}$
against $G^{h+1}=G\setminus\{G_{i}\vert1\leq i\leq h\}\setminus\{e\vert e\in f_{i,j_{i}^{*}},\,1\leq i\leq h\}$,
where $f_{i,j_{i}^{*}}$ is the flow already selected (i.e., $y_{i,j_{i}^{*}}$
is rounded to 1) for $\mathbb{F}_{i}$, and $G^{h+1}$ contains only
the edges corresponding to data items to be covered in future. Our
algorithm selects for $\mathbb{F}_{h+1}$ for the flow $f_{h+1,\, j_{h+1}^{*}}$
with $w(f_{i,h+1})+w(\mathbf{x}(G^{h+1}))$ maximized. The detailed
algorithm is shown in Algorithm \ref{alg:the1stderandomization}.
Why the algorithm removes the edges of $\{e\vert e\in f_{i,j_{i}^{*}},\,1\leq i\leq h\}$
from $G^{h+1}$ is that, if the edge corresponding to item $d_{l}$
is already in $\mathbb{F}_{i}:\, i\leq h$, the weight of the retrieve
sequence will not increase by covering any edge corresponding to the
same item $d_{l}$ for later processing.

\begin{algorithm}
\textbf{Input: }$\mathbb{F}=\{\mathbb{F}_{1},\dots,\mathbb{F}_{N}\}$,
where $\mathbb{F}_{i}=\{f_{i,\, j}\}$ is a collection of flows for
the $i$th segment of $\delta$A$\gamma$LWDR;

\textbf{Output:} $\mathbb{C}$, a solution to $\delta$A$\gamma$LWDR.
\begin{enumerate}
\item $\mathbb{C}:=\emptyset$;
\item Solve LP (\ref{eq:theoriginalLP}) by Karmarkar's algorithm \cite{schrijver1998theory},
and obtain an optimum solution $\mathbf{x}$;
\item \textbf{For} $i=1$ to $N$ \textbf{do }

\begin{enumerate}
\item \textbf{For} each $f_{i,j}\in\mathbb{F}_{i}$ \textbf{do}

\quad{}Set $z_{i,j}:=w(f_{i,j})+w(\mathbf{x}(G^{i}\setminus f_{i,j}))$;

\quad{}/{*} $z_{i,j}$ is the weight of $f_{i,j}$ plus the expected
weight sum of edge $e\in G^{i}\setminus f_{i,j}$ selected at probability
$x_{e}$. {*}/

\item Select $j_{i}^{*}$, such that $z_{i,j_{i}^{*}}\geq z_{i,j}$ for
every $f_{i,j}\in\mathbb{F}_{i}$;
\item $\mathbb{C}:=\mathbb{C}\cup\{f_{i,j_{i}^{*}}\}$
\end{enumerate}
\item Return $\mathbb{C}$.
\end{enumerate}
\protect\caption{\label{alg:the1stderandomization}Derandomization of Algorithm \ref{alg:theflows-randomized-algorithm}.}
\end{algorithm}

\begin{lem}
The ratio of Algorithm \ref{alg:the1stderandomization} is $1-\frac{1}{e}$.\end{lem}
\begin{IEEEproof}
Let $E(w(\mathbf{x}(G^{h})))$ denote the expectation weight sum of
edges $e\in G^{h}$ picked at probability $x_{e}$, i.e.

\begin{equation}
E\left(w\left(\mathbf{x}\left(G^{h}\right)\right)\right)=\sum_{i=1}^{j_{h}}y_{i,j_{i}}\cdot\left(w\left(f_{i,j_{i}}\right)+E\left(w\left(\mathbf{x}\left(G^{h}\setminus f_{i,j_{i}}\right)\right)\right)\right)\label{eq:defofEX}
\end{equation}

Then $E(w(\mathbf{x}(G^{1})))$ is equal to $E(w(SOL))$, the expectation
of the weight of the output of Algorithm \ref{alg:theflows-randomized-algorithm}.
So we need only to show $\sum_{i=1}^{N}w(f_{i,j_{i}^{*}})\geq E(w(\mathbf{x}(G^{1})))$,
provided that $E(w(SOL))\geq(1-\frac{1}{e})w(OPT)$ holds according
to Lemma \ref{lem:ratioOAflowLPadmits}.

First, for the last iteration, the algorithm picks $f_{N,\, j_{N}^{*}}$
with maximum $z_{N,j}:=w(f_{N,j})$ among all $j$s in $G^{N}$, so
the following inequality obviously holds:

\begin{equation}
w(f_{N,j_{N}^{*}})\geq E(w(\mathbf{x}(G^{N}))).\label{eq:N}
\end{equation}

Consider that we are selecting for $h$, then since $j_{h}^{*}$ is
chosen to attain maximum $z_{h,j_{h}^{*}}$, we have:

\begin{equation}
\begin{array}{ccc}
w\left(f_{h,j_{h}^{*}}\right)+E\left(w\left(\mathbf{x}\left(G^{h+1}\right)\right)\right) & \geq & \sum_{i=1}^{j_{h}}y_{i,j_{i}}\cdot\left(w\left(f_{i,j_{i}}\right)+E\left(w\left(\mathbf{x}\left(G^{h}\setminus f_{i,j_{i}}\right)\right)\right)\right)\end{array}\label{eq:main-2}
\end{equation}

Then, combining Inequality (\ref{eq:defofEX}) and (\ref{eq:main-2})
yields

\begin{equation}
w\left(f_{h,j_{h}^{*}}\right)+E\left(w\left(\mathbf{x}\left(G^{h+1}\right)\right)\right)\geq E\left(w\left(\mathbf{x}\left(G^{h}\right)\right)\right).\label{eq:deranfori}
\end{equation}

Summing up Inequality (\ref{eq:deranfori}) for every $1\leq h\leq N-1$,
and combining with Inequality (\ref{eq:N}), we have:

\[
\sum_{h=1}^{N}w\left(f_{h,j_{h}^{*}}\right)+\sum_{h=1}^{N-1}E\left(w\left(\mathbf{x}\left(G^{h+1}\right)\right)\right)\geq\sum_{h=1}^{N}E\left(w\left(\mathbf{x}\left(G^{h}\right)\right)\right).
\]

That is, $\sum_{i=1}^{N}w(f_{i,j_{i}^{*}})\geq E(w(\mathbf{x}(G^{1})))$.
This completes the proof.
\end{IEEEproof}

\section{An Improved Approximation Algorithm for General $\delta$A$\gamma$LWDR}

In this section, we shall show that Algorithm \ref{alg:theflows-randomized-algorithm}
can be extended to approximate general $\delta$A$\gamma$LWDR within
the same factor of $1-\frac{1}{e}$, without the occurrence assumption.
The key observation is that Algorithm \ref{alg:theflows-randomized-algorithm}
cannot produce a good approximation ratio for $\delta$A$\gamma$LWDR
since a fractional flow can contain two edges corresponding to an
identical data item. So the idea of the extension is to construct
an improved auxiliary graph in which different edges of an identical flow
is corresponding to distinct data items. Then, the algorithm is to
employ the collectively flow rounding method based on LP (\ref{eq:thedualLP})
given in this section against the improved auxiliary graph, and obtain
an approximation solution with the same ratio $1-\frac{1}{e}$.

\subsection{A Refined Construction and a Dual LP Formula}

Before giving the auxiliary graph where our algorithm can work correctly,
we would like first to give a refined construction of $G$, which
significantly decreases the number of edges of $G$, from $O(\vert V(G)\vert^{2})$
to $O(m\cdot\vert V(G)\vert)$, where $m$ is the number of the channels.
The key observation of the refined construction is that for a valid
retrieve sequence of LWDR, the constructed graph need only to contain
an $st$-path with all the edges, but not necessarily exactly the
same edges corresponding to the data items in the retrieve sequence.
Hence, we need only to construct a DAG satisfying Lemma \ref{lem:refinedconst}.
The key idea of the construction is to connect the head of an edge
to only one edge in every channel.

However, as analyzed later, LP (\ref{eq:theoriginalLP}) is not suitable
for $\delta$A$\gamma$LWDR with respect to the construction. Besides
the main part of $G$, it requires an additional virtual part to collaborate
a new LP relaxation. The full layout of the construction is in Algorithm
\ref{alg:RefinedConstruction}.

\begin{algorithm}
\textbf{Input: }An instance of $\delta$A$\gamma$LWDR;

\textbf{Output:} $G$.
\begin{enumerate}
\item Set $G:=\emptyset$, $G_{v}:=\emptyset$, $G_{r}:=\emptyset$; /{*}Initialization.
$G_{r}$ is for the main part that corresponding to $\delta$A$\gamma$LWDR;
$G_{v}$ the additional virtual part for a new LP relaxation.{*}/
\item For every item $d_{i}$:

\textbf{\quad{}}Add an edge set $E_{d_{i}}=\{e_{i,j,k}=(v_{i,j,k},\, w_{i,j,k})\vert d_{i}\mbox{ appears in channel \ensuremath{j}in the }k\mbox{th time slot}\}$
to $G_{r}$, where $w(e_{i,j,k})=w_{i}$;

\item Add two vertices $s$ and $p$ with weight-0 edges to $G_{r}$ as
below:

\begin{enumerate}
\item Edge $(s,\, v_{i,j,k})$ with minimum $k$ for every $j$;
\item Edge $(w_{i,j,k},\, p)$ with maximum $k$ for every $j$;;
\end{enumerate}
\item \textbf{For} $j=1$ to $m$ \textbf{do}

\textbf{\quad{}For} $k=1$ to $T$ \textbf{do \quad{}}/{*} Add edges
for the relationship between the items. {*}/

\textbf{\quad{}\quad{}If} edge $(v{}_{i,j,k},\, w_{i,j,k})$ exists
\textbf{then}

\textbf{\quad{}\quad{}\quad{}}Add weight-0 edge $(w{}_{i,j,k},\, v_{i',j,k'})$
to $G_{r}$ with minimum $k'>k$;

\textbf{\quad{}\quad{}\quad{}}Add weight-0 edge $(w_{i,j,k},\, v_{i',j',k'})$
to $G_{r}$ with minimum $k'>k+1$ for every $j'\neq j$;

\textbf{\quad{}\quad{}EndIf}

/{*}For each item add an edge from $v{}_{i,j,k}$ to the tail of the
edge corresponding to the nearest item that could be retrieved conflict-freely
afterward.{*}/

\item \textbf{For} every item $d_{i}$ \textbf{do}

\textbf{\quad{}}Add edges $e_{i}=(u_{T+i},v_{T+i})$ and $e'_{i}=(u'_{T+i},v'_{T+i})$,
as well as the edges $e_{i}=(u_{T+i},v'_{T+i})$ and $e'_{i}=(u'_{T+i},v_{T+i})$,
to $G_{v}$ where $c(e_{i})=1$ and $w(e_{i})=w(d_{i})$, other edges
are all with cost 0.

/{*} The construction of $G_{v}$. Note that $v_{T+i}=u_{T+i+1}$,
$v'_{T+i}=u'_{T+i+1}$, and $u_{T+n}=u'_{T+n}=t$.{*}/

\item Add edges $e(p,u_{T+1})$ and $e(p,u'_{T+1})$ to $G_{v}$;

/{*} Add the connection between $G_{r}$ and $G_{v}$. {*}/

\item Return $G:=G_{r}\cup G_{v}$.
\end{enumerate}
\protect\caption{\label{alg:RefinedConstruction}Construction of Auxiliary Graph for
$\delta$A$\gamma$LWDR. }
\end{algorithm}

\begin{figure}
\includegraphics{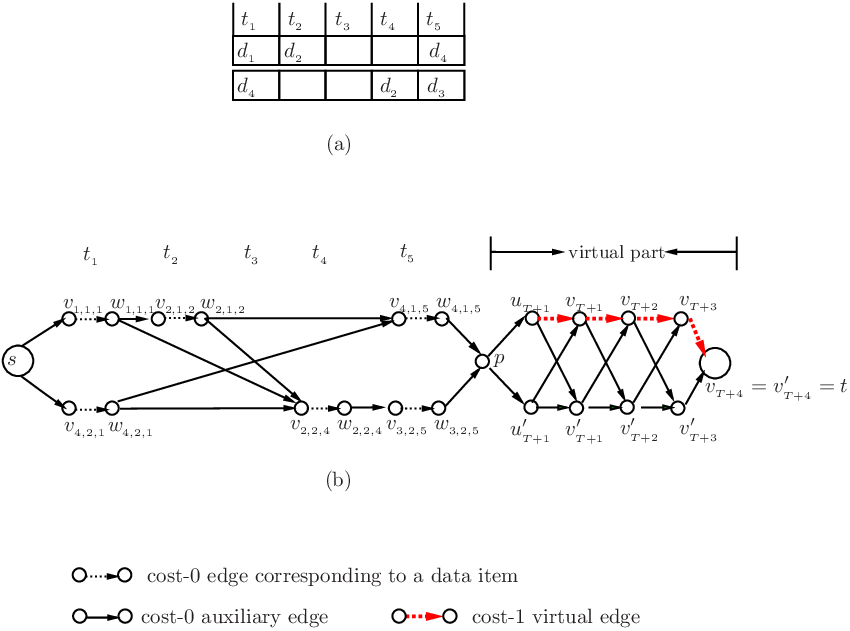}

\protect\caption{\label{fig:A-Refined-Construction}A refined construction of DAG:
(a) An instance of LWDR; (b) The refined corresponding graph. }
\end{figure}

\begin{lem}
\label{lem:refinedconst}$S$ is a valid retrieve sequence for $\delta$A$\gamma$LWDR
if and only if in the constructed graph of Algorithm \ref{alg:RefinedConstruction}
there exist $\delta$ disjoint \textbf{$st$-}paths whose corresponding
retrieve sequences contain every item of $S$.  \end{lem}
\begin{IEEEproof}
We shall only show the lemma holds for the case $\delta=1$, since
the case for general $\delta$ is similar.

For the ``only if'' direction, let $S=o_{i_{1},j_{1},k_{1}},o_{i_{2},j_{2},k_{2}},\dots o_{i_{l},j_{l},k_{l}},o_{i_{l+1},j_{l+1},k_{l+1}},\dots$,
$k_{l+1}>k_{l}$ for any $l$, be a retrieve sequence for an instance
LWDR, where $o_{i_{l},j_{l},k_{l}}$ is data item $d_{i_{l}}$ retrieved
in channel $j_{l}$ and time slot $k_{l}$. If $o_{i_{l+1},j_{l+1},k_{l+1}}$
can be retrieved after $o_{i_{l},j_{l},k_{l}}$, then the two data
items must be conflict-free. That is, $o_{i_{l+1},j_{l+1},k_{l+1}}$
and $o_{i_{l},j_{l},k_{l}}$ are either (1) in the same channel, i.e.
$j_{l_{1+1}}=j_{l}$; or (2) in different channels and $k_{l+1}-k_{l}\geq2$.
According to the construction of $G$, there will be an edge leaving
$w_{i_{l},j_{l},k_{l}}$, the head of the edge corresponding to item
$o_{i_{l},j_{l},k_{l}}$, and entering $v{}_{i_{l+1},j_{l+1},k'}$,
the tail of the edge corresponding to a conflict-free data item $o_{i_{l+1},j_{l+1},k'}$
for the minimum $k'>k_{l}$. Then according to the construction again,
there exists a path from $o_{i_{l+1},j_{l+1},k'}$ to $o_{i_{l+1},j_{l+1},k_{l+1}}$.
That is, every $o_{i_{l+1},j_{l+1},k_{l+1}}$ is reachable from $v_{i_{l},j_{l},k_{l}}$.
Therefore, $P=s,v_{i_{1},j_{1},k_{1}},w_{i_{1},j_{1},k_{1}},v_{i_{2},j_{2},k_{2}},w{}_{i_{2},j_{2},k_{2}},\dots,t$
is an $st$ path in $G$.

For the ``if'' direction, assume that there exists a path $P=s,v_{i_{1},j_{1},k_{1}},w_{i_{1},j_{1},k_{1}},v_{i_{2},j_{2},k_{2}},w_{i_{2},j_{2},k_{2}},\dots,$
$v_{i_{l},j_{l},k_{l}},w_{i_{l},j_{l},k_{l}},v_{i_{l+1},j_{l+1},k_{l+1}},w_{i_{l+1},j_{l+1},k_{l+1}},\dots,t$,
$k_{l+1}>k_{l}$, in $G$. According to the construction, for any
edge $(w_{i_{l},j_{l},k_{l}},v{}_{i_{l+1},j_{l+1},k_{l+1}})$, $k_{l+1}-k_{l}>1$
holds or $j_{l}=j_{l+1}$ and $k_{l+1}>k_{l}$ both hold. That is,
the items retrieved in time slot $k_{l}$ and $k_{l+1}$, say $o_{i_{l},j_{l},k_{l}}$
and $o_{i_{l+1},j_{l+1},k_{l+1}}$ can be retrieved conflict-freely.
So $o_{i_{1},j_{1},k_{1}},o_{i_{2},j_{2},k_{2}},\dots$ is a valid
retrieve sequence. These completes the proof.
\end{IEEEproof}
It is easy to see that the DAG $G$ resulting from Algorithm \ref{alg:RefinedConstruction}
has a much smaller size compared to the previous construction as in
Algorithm \ref{alg:1Construction-of-Auxiliary}, as stated below:
\begin{lem}
\label{lem:sizeofrefined}In the constructed graph $G$, there exist
at most $O(m\cdot T)$ vertices, and at most $O(m^{2}T)$ edges.\end{lem}
\begin{IEEEproof}
Clearly, we have $\vert V(G)\vert=O(m\cdot T)$. That is because we
add an edge for each occurrence of the data items in the construction,
and the number of the occurrences is at most $O(m\cdot T)$. Then
since there exist at most $m$ edges leaving a vertex, $E(G)\leq m\cdot\vert V(G)\vert=O(m^{2}T)$.
\end{IEEEproof}
We now argue that LP (\ref{eq:theoriginalLP}) is no longer a relaxation
for $\delta$A$\gamma$LWDR with respect to the above refined construction.
Because for a solution of $\delta$A$\gamma$LWDR, there might exist
no $st$-path with exactly the edges corresponding to the retrieved
sequence, i.e. because a path in $G$ might contain two edges that
corresponding to an identical data item (See figure \ref{fig:A-Refined-Construction}
for an example: The path, corresponding to the retrieve sequence containing
$d_{1},d_{2},d_{3}$, is forced to go through both edges $(v_{2,1,2},w_{2,1,2})$
and $(v_{2,2,4},w_{2,2,4})$, which are corresponding to the an identical
item). But according to Inequality (\ref{eq:keyconstr}), $\sum_{e\in E_{d_{i}}}x_{e}\leq1$
should hold. Therefore, it remains to give a new LP formula for $\delta$A$\gamma$LWDR
with respect to $G$ resulted from the refined construction. To do
so, we first add a virtual part to $G$, and then give an LP formula,
which allows a path in $G$ to contain multiple edges corresponding
to an identical item. Then the LP formula is as below:

\begin{eqnarray}
min &  & \sum_{i=N+1}^{N+n}c(e_{i})\cdot w(e_{i})\cdotp x_{e_{i}}\label{eq:thedualLP}\\
s.t. &  & \sum_{e\in\delta^{+}(v)}x_{e}-\sum_{e\in\delta^{-}(v)}x_{e}=\left\{ \begin{array}{cc}
0 & \mbox{ }\quad\forall v\in V\setminus\{s,\, t\}\\
\delta & v=s
\end{array}\right.\label{eq:refinedLPflow}\\
 &  & \sum_{j,k}x_{e_{i,j,k}}\geq1\,\quad\quad\quad\quad\quad\quad\quad\quad\mbox{ }1\leq i\leq n\\
 &  & 0\leq x_{e}\leq1\,\quad\quad\quad\quad\quad\quad\quad\quad\quad\mbox{ }\forall e\in G\label{eq:x_ebetween01}
\end{eqnarray}
where $x_{e}=1$ iff $e$ is selected and $x_{e}=0$ otherwise. If
$x_{e}\in\{0,\,1\}$, then from Lemma \ref{lem:refinedconst}, the
above formula becomes an integral programming (IP) formula for $\delta$A$\gamma$LWDR.
The intuitive explanation of the above LP is as below: Let $W=\sum_{i=1}^{n}w_{i}$,
let $c_{IP}$ be the minimum of the objective function of the IP corresponding
to LP (\ref{eq:LP}), and let $w_{OPT}$ be the weight of an optimal
solution to the corresponding $\delta$A$\gamma$LWDR instance. Then,
we have
\begin{equation}
W=c_{IP}+w_{OPT}.\label{eq:w-1}
\end{equation}
Because $W$ is fixed, it is identical either to maximize $w_{OPT}$
or to minimize $c_{IP}$.

Then based on the new LP formula, Algorithm \ref{alg:theflows-randomized-algorithm}
can be immediately adopted to solve $\delta$A$\gamma$LWDR. Following
the same line of Lemma \ref{lem:ratioOAflowLPadmits} and \ref{lem:flowOAtime},
and then Lemma \ref{lem:sizeofrefined} on the sized of $G$, we have
the following Theorem:
\begin{thm}
Under the occurrence assumption, $\delta$A$\gamma$LWDR admits an
algorithm with ratio $1-\frac{1}{e}$ and a time complexity $O(m^{7}T^{3.5}L)$.
\end{thm}

\subsection{Construction of the Improved Auxiliary Graph}

\begin{figure}
\includegraphics{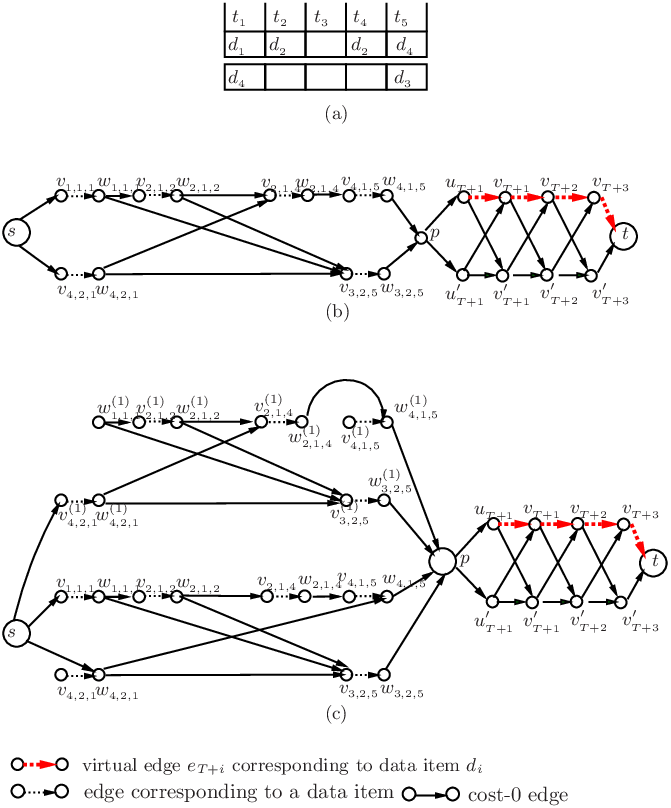}

\protect\caption{\label{fig:oneoccuranceitem}An construction of the improved DAG.}
\end{figure}

This subsection will further improve the auxiliary graph output by
Algorithm \ref{alg:RefinedConstruction}, such that in the improved
auxiliary graph any path in a segment will not go through two edges
corresponding to an identical data item. Thus, the improved auxiliary
graph can be considered as a graph satisfying the occurrence assumption,
and hence Algorithm \ref{alg:theflows-randomized-algorithm} can be
employed to solve the $\delta$A$\gamma$LWDR problem accordingly.
The key idea of the improved construction is to find and eliminate
every pair of edges which correspond to an identical data item and
appear in a common path within a segment. The detailed construction
of the improved auxiliary graph is in Algorithm \ref{alg:enhanced-construction},
where w.l.o.g. we assume the $\delta$A$\gamma$LWDR instance has
only one segment, since the case for multiple segments is similar.
An example of execution of the algorithm is depicted in Figure \ref{fig:oneoccuranceitem}.

\begin{algorithm}
\textbf{Input: }A refined auxiliary graph $G$ (for $\delta$A$\gamma$LWDR
with one segment) output by Algorithm \ref{alg:RefinedConstruction};

\textbf{Output: }A graph $G'$, in which no flow exists containing
two edges corresponding to an identical data item.
\begin{enumerate}
\item $G':=G$;
\item \textbf{For} $h=1$ to $T$ \textbf{do}

\begin{enumerate}
\item \textbf{For} each edge $e=(v,\, w)\in G[t_{h+1},\, p]$ \textbf{do
}/{*}\textbf{$G[t_{h+1},\, p]$ }is the subgraph of $G$ from time
slot $t_{h+1}$ to $p$.{*}/

\quad{}Add a corresponding edge duplicating $e$, say $e^{(k)}=(v^{(k)},\, w^{(k)})$,
to $G^{'}$, assuming it is the $k$th time duplicating $e$;

\quad{}/{*}Duplicate $G[t_{h+1},\, p]$ by duplicating every edge
therein. {*}/

\item \textbf{For} $j=1$ to $m$ \textbf{do}

\quad{}\textbf{while} there exist both edge $(v_{i,j,h},w_{i,j,h})$
and edge $(v_{i,j',h'},w{}_{i,j',h'})$ with $h'>h$, such that $v_{i,j',h'}$
is reachable from $w_{i,j,h}$ \textbf{do}

\quad{}/{*}There exist a path containing 2 edges both corresponding
to item $d_{i}$.{*}/

\quad{}\quad{}(i) Replace each edge ended at $v_{i,j,h}$, say $e(v,v_{i,j,h})$,
with $e(v,w{}_{i,j,h})$ in $G'$;

\quad{}\quad{}(ii) Replace each edge ended at $v_{i,j',h'}^{(k)}$,
say $e(v,v_{i,j',h'}^{(k)})$, with $e(v,w{}_{i,j',h'}^{(k)})$ in
$G'$;

\end{enumerate}

\textbf{Endfor}

\item Return $G'$.
\end{enumerate}
\protect\caption{\label{alg:enhanced-construction}The construction of an improved
auxiliary graph.}
\end{algorithm}

\begin{lem}
\label{lem:timeofalgf}In runtime $O(2^{\gamma}\vert E(G)\vert)$,
Algorithm \ref{alg:enhanced-construction} outputs a graph $G'$ with
at most $O(2^{\gamma}\vert E(G)\vert)$ edges, where $\gamma$ is
the maximum length of a segment of $\delta$A$\gamma$LWDR. $G'$
has a size, i.e. $\max\{E(G'),V(G')\}$, not larger than $2^{\gamma}\cdot\vert E(G)\vert$,
and every path therein contains at most one edge of $E_{d_{i}}$ for
each item $d_{i}$ within a segment. \end{lem}
\begin{IEEEproof}
For the time complexity, Step 2 of Algorithm \ref{alg:enhanced-construction}
repeats at most $O(\gamma)$ times, each of which at most doubles
the size of $G'$. Since $G'$ is initially $G$, $\vert E(G')\vert\leq2^{h-1}\cdot\vert E(G)\vert$
holds at the beginning of the $h$th iteration. In this iteration,
it takes $O(E(G'))$ time to duplicate the edges (in Step 2(a)) and
takes $O(E(G'))$ time to check for $(v_{i,j,k},w_{i,j,k})$ for every
$j$ that whether there exists a path containing 2 edges both corresponding
to an identical item (in Step 2(c)). So the total runtime of the algorithm
is $\sum_{h=1}^{\gamma}O(2^{h-1}\cdot\vert E(G)\vert)=O(2^{\gamma}\cdot\vert E(G)\vert)$.
In addition, we also have $E(G')=O(2^{\gamma}\cdot\vert E(G)\vert)$.

According to Algorithm \ref{alg:enhanced-construction}, the $h$th
iteration guarantees that no path can contain both edge $(v_{i,j,h},w{}_{i,j,h})$
and another edge corresponding to $d_{i}$. Therefore, when Algorithm
\ref{alg:enhanced-construction} terminates, all paths containing
2 edges corresponding to an identical item are eliminated. This completes
the proof.
\end{IEEEproof}

\begin{figure}
\includegraphics[scale=0.8]{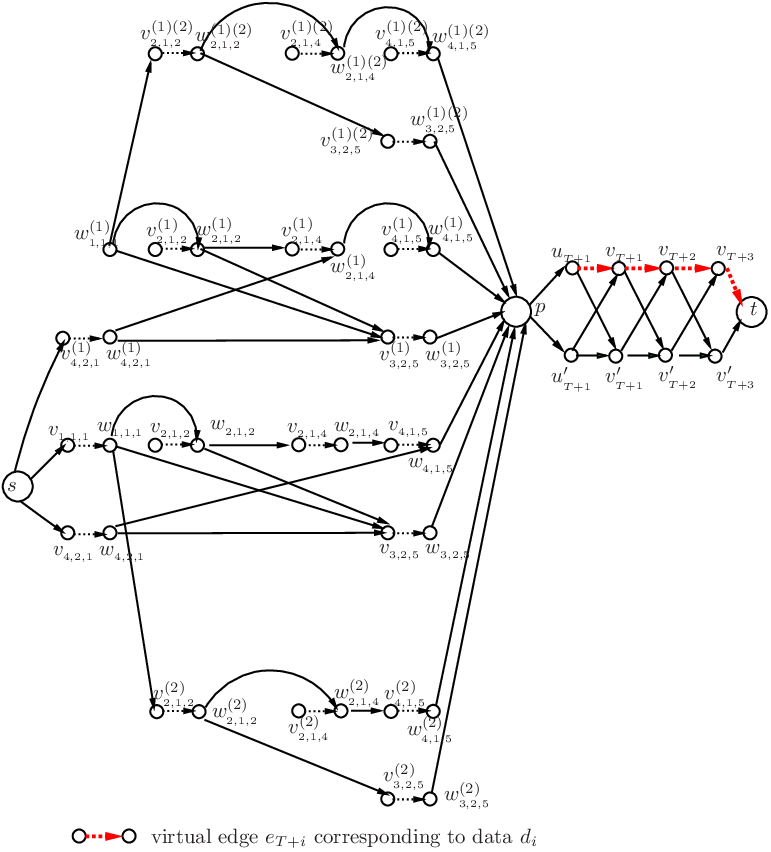}

\protect\caption{\label{fig:oneoccuranceitem-1}An construction of the improved DAG
(continued)}
\end{figure}

The correctness of Algorithm \ref{alg:enhanced-construction} can
be immediately obtained from the following lemma:
\begin{lem}
\label{lem:finalgra}There exists $\delta$-disjoint $st$-path $P_{1},\,\dots,\, P_{\delta}$
in $G$ if and only if there exists in $G'$ $\delta$ disjoint \textbf{$st$-}paths
$Q_{1},\dots,Q_{\delta}$, such that their corresponding retrieve
sequences\textbf{ }contain identical data items.\end{lem}
\begin{IEEEproof}
Let $e_{i,j,k}$ and $e_{i',j',k'}$ be two edges in $G$ output by
Algorithm \ref{alg:RefinedConstruction}, where $e_{i,j,k}$ means
that data item $d_{i}$ appears in channel $j$ at time slot $k$.
According to the construction of $G'$ as in Algorithm \ref{alg:enhanced-construction},
if $e_{i,j,k}$ and $e_{i',j',k'}$ connected in $G$, then any duplication
of $e_{i,j,k}$ and that of $e_{i',j',k'}$ are connected in $G'$
when $i\neq i'$. Let $P$ containing edges $\{e_{i_{1},j_{1},k_{1}},\,\dots,\, e_{i_{h},j_{h},k_{h}}\}$,
$k_{l}<k_{l+1}$, be a path in $G$, where e $k_{l}$ is the minimum
time slot where $d_{i_{l}}$'s corresponding edges appear on the path
$P$. Then there must exist a path $Q$ containing $\{e_{i_{1},j_{1},k_{1}}^{(*)},\,\dots,\, e_{i_{h},j_{h},k_{h}}^{(*)}\}$
in $G'$, where $e_{i_{1},j_{1},k_{1}}^{(*)}$ means any duplication
of $e_{i_{1},j_{1},k_{1}}$. Therefore, there exists $Q$ in $G'$
which retrieves the same data items $\{d_{i_{1}},\,\dots,\, d_{i_{h}}\}$
as $P$ does. Similarly and conversely, we can construct a path $P$
in $G$ from Q in $G'$, such that $P$ and $Q$ contain identical
data items. This completes the proof. \end{IEEEproof}
\begin{thm}
\label{thr:kagammaLWDR} $\delta$A$\gamma$LWDR admits an approximation
algorithm with a ratio $1-\frac{1}{e}$ and a time complexity $O(2^{\gamma}m^{7}T^{3.5}L)$,
where $m$ is the number of channels, $T$ is the number of time slots,
and $L$ is the maximum length of the input.\end{thm}
\begin{IEEEproof}
The ratio can be easily obtained by combining Lemma \ref{lem:finalgra}
and \ref{lem:ratioOAflowLPadmits}. For the time complexity, it takes
$O(\vert E(G)\vert^{3.5}L)$ time to solve LP (\ref{eq:thedualLP}),
since the number of the constraints is $O(\vert E(G)\vert)$ \cite{korte2002combinatorial}.
Then by Lemma \ref{lem:timeofalgf}, the runtime $O(2^{\gamma}m^{7}T^{3.5}L)$
follows.
\end{IEEEproof}
For any given $\epsilon>0$, by setting $\gamma=\left\lfloor \frac{1}{\epsilon}\right\rfloor $
when transforming from $\delta$A$\gamma$LWDR to $\delta$ALWDR,
and then combining Theorem \ref{thr:kagammaLWDR} and Proposition
\ref{prop:simpobsbetw-gamma}, we have:
\begin{thm}
\label{thm:finalratioforkLWDR}For any fixed $\epsilon>0$, $\delta$ALWDR
admits an approximation algorithm with a ratio $1-\frac{1}{e}-\epsilon$
and a runtime $O(2^{\frac{1}{\epsilon}}\frac{1}{\epsilon}m^{7}T^{3.5}L)$.
\end{thm}
As a by-production, it can be shown that $\delta$ALWDR is fixed parameter
tractable with respect to the parameter ``$B$'':
\begin{cor}
$\delta$ALWDR admits an exact algorithm with a time complexity $O(2^{B}m^{7}T^{3.5}L)$,
where $B$ is the number of time slots containing data items which
has occurrences in subsequent time slots.\end{cor}
\begin{IEEEproof}
For any instance of $\delta$ALWDR, the exact algorithm is first to
run Algorithm \ref{alg:enhanced-construction} against the instance,
and then to solve the according LP \ref{eq:thedualLP} to get a basic
optimum solution. Then since $\vert E(G')\vert=O(2^{B}\cdot\vert E(G)\vert)=O(2^{B}m^{2}T)$
from Lemma \ref{lem:timeofalgf}, the time complexity of the exact
algorithm would be $O(2^{B}m^{7}T^{3.5}L)$ in worst case. For the
correctness, from Lemma \ref{lem:timeofalgf}, in $G'$ there exists
no path containing two edges corresponding to one identical item.
That is, the task remains only to compute a set of $\delta$-disjoint
longest paths in $G'$, which is corresponding to an optimum solution
for $\delta$ALWDR. Following the same line of the proof of Theorem
\ref{Thr:LWDR-is-polynomial-1}, the task can be done in polynomial
time $O(2^{B}m^{7}T^{3.5}L)$.
\end{IEEEproof}

\section{${\cal NP}$-Completeness of $\delta$A$\gamma$LWDR }

In this section, we show that $\delta$A$\gamma$LWDR remains ${\cal NP}$-complete
for $\gamma=2$, by giving a reduction from the 3-dimensional perfect
matching (3DM) problem, which is known ${\cal NP}$-complete \cite{garey1979computer}.
Moreover, the ${\cal NP}$-completeness remains true even when there
are only three channels, every item is with the same weight and appears
at most 3 times.

For a given positive integer $K$ and a set $T\subseteq X\times Y\times Z$
where $X$, $Y$ and $Z$ are disjoint and $|X|=|Y|=|Z|$, 3DM is
to decide whether there exists a perfect matching for $T$, i.e.,
a subset $M\subseteq T$ with $|M|=|X|$, such that no elements in
$M$ agree in any coordinate.
\begin{thm}
\label{thm:kagammaLWDR-is-NP-complete,} $\delta$A$\gamma$LWDR is
${\cal NP}$-complete even when $\gamma=2$.
\end{thm}
Since $\delta$A$\gamma$LWDR is evidently in ${\cal NP}$, we need
only to give the reduction from 3DM to the decision form of $\delta$A$\gamma$LWDR:
Given a positive integer $K$ and $D$, a set of items of equal weight
1 broadcast in the channels in a time interval, does there exist a
retrieve sequence of $D$ with total weight not less than $K$?

\begin{figure}
\includegraphics{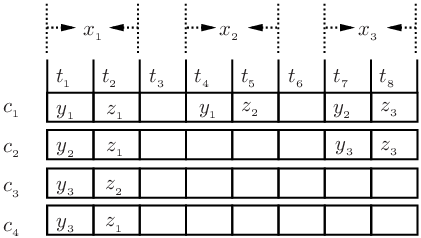}

\protect\caption{\label{fig:nproof}An instance of $\delta$A$2$LWDR corresponding
to the following instance of 3DM: $T=\{\left\langle x_{1},\, y_{1},\, z_{1}\right\rangle ,\,\left\langle x_{1},\, y_{2},\, z_{1}\right\rangle ,\,\left\langle x_{1},\, y_{3},\, z_{2}\right\rangle ,\,\left\langle x_{1},\, y_{3},\, z_{1}\right\rangle ,\,\left\langle x_{2},\, y_{1},\, z_{2}\right\rangle ,\,\left\langle x_{3},\, y_{2},\, z_{3}\right\rangle ,\,\left\langle x_{3},\, y_{3},\, z_{3}\right\rangle \}$.}
\end{figure}

For an instance of decision 3DM, the construction of the corresponding
$\delta$A$\gamma$LWDR instance is simply as below (An example of
the construction is depicted in Figure \ref{fig:nproof}):
\begin{enumerate}
\item For each element of $X$, say $x_{i}$, add 3 time slots $t_{3i+1}$,
$t_{3i+2}$, $t_{3i+3}$ to the time interval, which are initially
empty time slots;
\item For each element of $T$, say $\left\langle x_{i},\, y_{j},\, z_{k}\right\rangle $
where $\left\langle x_{i},\, y_{j},\, z_{k}\right\rangle $ is the
$l$th occurrence of $x_{i}$, broadcast two items $y_{j}$ and $z_{k}$ in
time slots $t_{3i+1}$ and $t_{3i+2}$ of channel $l$, respectively.
\end{enumerate}

\begin{IEEEproof}
Note that in the instance of $\delta$A$\gamma$LWDR constructed as
above, clearly $\gamma=2$. We need only to show that an instance
of 3DM is feasible if and only the corresponding $\delta$A$\gamma$LWDR
is feasible for $K=2\vert X\vert$.

Firstly, assume that for the given 3DM instance there exists a perfect
matching, say \[M=\left\{ \left\langle x_{1},\, y_{1},\, z_{1}\right\rangle ,\,\left\langle x_{2},\, y_{2},\, z_{2}\right\rangle ,\,\dots,\,\left\langle x_{|X|},\, y_{|X|},\, z_{|X|}\right\rangle \right\} \],
such that no elements in $M$ agree in any coordinate, i.e. for any
$l\neq l'$, $\{x_{l'},\, y_{l'},\, z_{l'}\}\cap\{x_{l},\, y_{l},\, z_{l}\}=\emptyset$
holds. Then for the constructed $\delta$A$\gamma$LWDR instance,
clearly $P=y_{1},\, z_{1},\, y_{2},\, z_{2},\,\dots,\, y_{|X|},\, z_{|X|}$
is a feasible solution with $\vert P\vert=2|X|$.

Conversely, assume that $P=y_{1},\, z_{1},\, y_{2},\, z_{2},\,\dots,\, y_{|X|},\, z_{|X|}$
is a feasible solution with $\vert P\vert\geq2|X|$ to $\delta$A$\gamma$LWDR.
Then each data item in $P$ is distinct, because for every $i\in\{1,\,\dots,\,|X|\}$,
$y_{i}$ and $z_{i}$ must contribute weight 2 to the total weight.
That is, for any $l\neq l'$, $\{y_{l'},\, z_{l'}\}\cap\{y_{l},\, z_{l}\}=\emptyset$.
Therefore, $M=\left\{ \left\langle x_{1},\, y_{h_{1}},\, z_{k_{1}}\right\rangle ,\,\left\langle x_{2},\, y_{h_{2}},\, z_{k_{2}}\right\rangle ,\,\dots,\,\left\langle x_{|X|},\, y_{h_{|X|}},\, z_{k_{|X|}}\right\rangle \right\} $
is a perfect matching for the given 3DM instance. This completes the
proof.
\end{IEEEproof}
Further, the ${\cal NP}$-completeness remains true even for a very
special case of $\delta$A$\gamma$LWDR :
\begin{cor}
$\delta$A$\gamma$LWDR is ${\cal NP}$-complete, even when $\gamma=2$,
only three channels exist, and every item is with equal weight and
appears at most 3 times in the time interval.\end{cor}
\begin{IEEEproof}
It is known that 3DM is ${\cal NP}$-complete even if the number of
occurrences of any element in $X$, $Y$ or $Z$ is bounded by ``3''
\cite{kann1991maximum}. According to the transformation, because
of the bound ``3'' on element occurrences of 3DM, both the number
of channels and the occurrences of any data item can be bounded by
3.
\end{IEEEproof}

\section{Performance Evaluation}

In this section, we show performance evaluation of our algorithms
(RFA, Algorithm \ref{alg:theflows-randomized-algorithm}) by experiments,
and compare our algorithms with the approximation algorithm adopting
maximum weight matching (MM, as in \cite{Infocom12LuEfficient}) when
$\delta=1$. Experimental results comparing our algorithm with an
exactly algorithm (EA, based on integral linear programming (ILP))
for $\delta=2$ is also given in the appendix. We implement the algorithms
using python 2.7, on a PC with Mac OS X Yosemite, 1.4 GHz Intel Core
i5 processor, and 8GB 1600MHz DDR3 memory. Other than our proposed
algorithm, we also implement the maximum matching heuristic, and the
exact algorithm (EA). Our implementation uses the \emph{networkx}
library to construct both the auxiliary graphs of RFA and MM, the
interior-point method of the \emph{GLPK} library to solve LPs and
the simplex method of \emph{GLPK }to\emph{ }solve ILPs.

\subsection{Methodology}

To evaluate our algorithms, we simulate push-based broadcast programs.
We denote the number of down-link channels by $m$, the total number
of time slots by $T$, the total number of broadcasting data items
by $N$, and the number of the packets in a request by $n$. In our
experiments, $m$ is set in the range of $[2,\,16]$, $T$ in $[100,\,500]$,
$N$ in $[100,\,5000]$, and $n$ in $[100,\,500]$. We assume that
all the channels are with uniform bandwidth, and all broadcast data
items are with the same size. Besides, for RFA, we set the value of
$\gamma$ in $[3,\,50]$. In our experiments, for a given the skewed
parameter $\theta$, we also assume the access probability of a data
item $d_{i}$ for a request follows the Zipf distribution :

\[
p_{i}=\frac{i^{-\theta}}{\sum_{i=1}^{N}i^{-\theta}}.
\]

We use average download percentage (ADP) as the performance metric,
and simulate 10,000 requests to get ADP for each experiment.

\subsection{Experimental Results }

\begin{figure}
\subfloat[$m=2$, n=400, $\gamma=2,\,\dots,\,22$.]{\includegraphics[width=0.5\columnwidth]{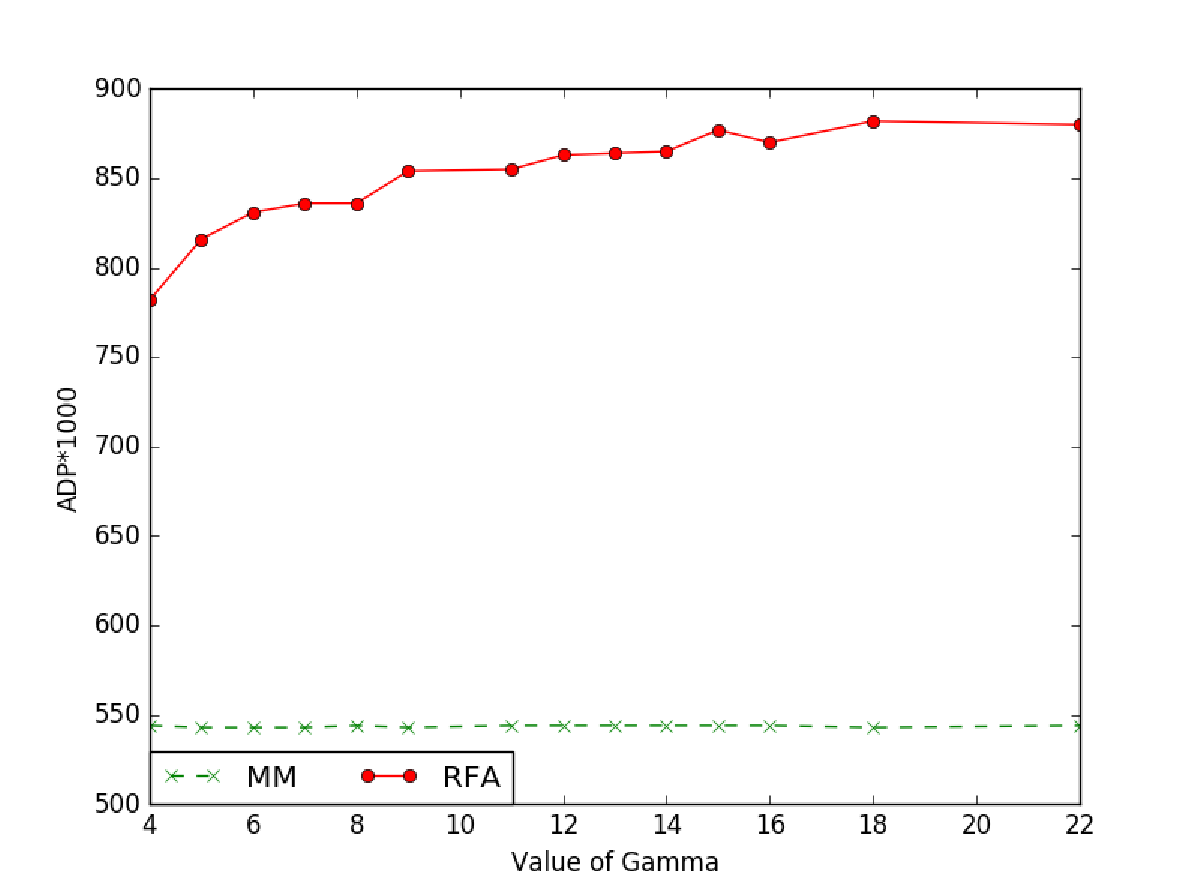}

} \subfloat[$m=8$, n=1600, $\gamma=2,\,\dots,\,22$.]{\includegraphics[width=0.5\columnwidth]{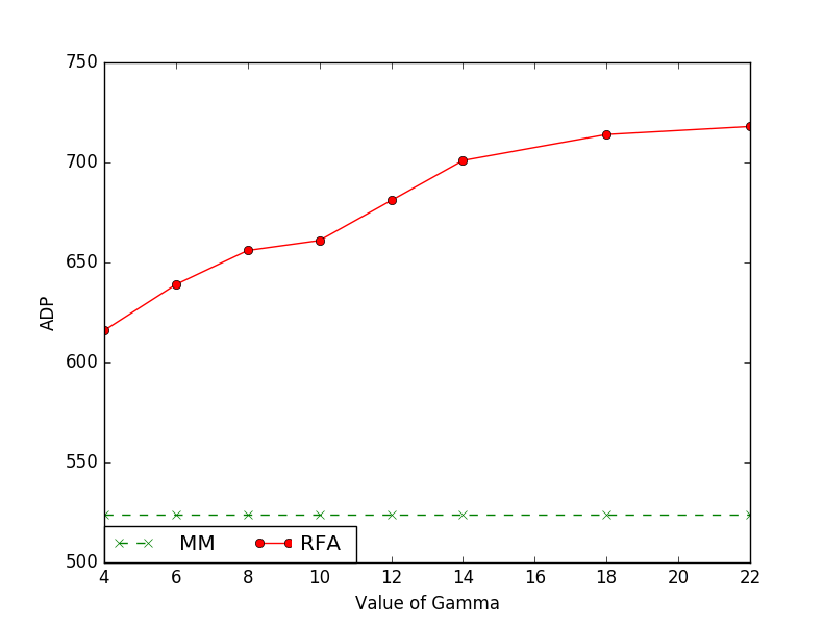}

}

\protect\caption{\label{fig:gamma}LWDR, $T=200$, $p=150$. }
\end{figure}

The simulation results comparing RFA and MM are depicted in Figure
\ref{fig:gamma}. In all the experiments, the ADP of RFA always achieves
better performance than that of MM: by about 20-32 percent in Figure
\ref{fig:gamma} (a), and by about 9-20 percents in Figure \ref{fig:gamma}
(b). These results are actually better than our analysis. This phenomenon
is reasonable, because our algorithm is based on rounding a fractional
solution of the LP relaxation of the problem. In many cases, rounding
such an LP optimal solution could result in actually a much better
ADP than the worst case as analyzed. Comparing Figure \ref{fig:gamma}
(a) and Figure \ref{fig:gamma} (b), we find the ADP of RFA decreases
when $m$ increases. That is because the problem becomes more complicated
when $m$ grows, and hence the quality of the solution decreases accordingly.
However, RFA still significantly outperforms MM when $m=8$, although
the gap between them decreases. For the relationship between ADP and
$\gamma$, as shown in Figure \ref{fig:gamma} (a) and (b), the ADP
of RFA increases when $\gamma$ increases. This is consistent with
our analysis of Algorithm \ref{alg:theflows-randomized-algorithm}:
as $\gamma$ increases, $\epsilon=\frac{1}{\gamma}$ then decreases,
and hence the approximation ratio $1-\frac{1}{e}-\epsilon$ increases.
In particular, as we could see in Figure \ref{fig:gamma}(a), ADP
increases significantly from roughly 78 to 86 percents, when $\gamma$
increases significantly from 4 to 12; when $\gamma>12$, the impact
of $\gamma$ over ADP grows inefficiently, i.e. ADP barely increases
when $\gamma$ increases. Then ADP attains the maximum at about 88
percents at $\gamma=18$, and then ADP decreases when $\gamma>18$
grows. While $m$ is 8 instead of 2 in Figure \ref{fig:gamma}(b),
the weight increasing upon $\gamma$'s increment remains efficient
until $\gamma>18$. Thus, for a more complicated instance, the increment
of $\gamma$ benefits for a larger range.

The simulation results on the runtime of the algorithms are as given
in Table \ref{tab:Runtime-analysis.}, in which three parameters are
considered: $T$, $m$ and $\gamma$, since according to our theoretical
analysis in the previous section, these parameters are the factors
that affect the runtime of the algorithm. In all the runtime experiments,
the parameters are set $\delta=2$, $\gamma=10$ and $m=4$, where
$\delta=2$ is set, since most of the mobile devices in not-long future
is likely to have two antennae, provided that the current mobile devices
are mostly with only one antenna. Then $m=4$ is a reasonable number
of channels for $\delta=2$, while $\gamma=10$ is likely the value
of $\gamma$ with best performance time ratio. Comparing the runtime
of RFA and EA, we can see that the runtime of RFA is significantly
lower than EA, particularly when the problem size is large, i.e. the
involved $T$ is with size 500. Better still, the gap between them
grows when the size of the problem increases. It is worth to note
that, the time of EA is mostly the time of solving ILP, while the
solver of ILP used in EA is the \emph{GLPK} ILP solver, a mature package
that is almost perfectly implemented. That is, with better implementation
of RFA, RFA could have a runtime over EA even better than as in the
table. On the other hand, we note that the runtime of RFA is higher
than MM when the problem size is large. The high runtime mainly comes
from the cost of solving the LP formula. When $m=4$, $T=1500$ and
$\gamma=10$, the algorithm is to solve an LP of size about 20,000.
However, we argue the algorithm has practical values. Firstly, the
algorithm can be implemented in a better way. The current implementation
is based on Python, so the runtime could be better if the algorithm
is implemented in other languages such as C or C++ that is known with
a better performance. Besides, the open source library \emph{GLPK}
that we used to solve LP, takes almost 500s to solve an LP of size
20, 000. Note that this runtime could be improved if using some other
commercial libraries such as \emph{Gurobi optimizer} which as claimed
is more efficient than \emph{GLPK} and can solve LP with up to millions
of variables in a reasonable time. Last but not least, there are a
lot of efficient approximation algorithms that solve LP even faster,
adopting which could further improve the runtime of our algorithms.
Secondly, even with our simple implementation, the average runtime
of our algorithm is 500ms for LWDR with the number of request data
items of 100 and a channel number of 8. Therefore, Algorithm \ref{alg:theflows-randomized-algorithm}
has the potential to be applied in real networks.

\begin{table*}
\protect\caption{\label{tab:Runtime-analysis.}Runtime analysis.}

\subfloat{\centering{}%
\begin{tabular}{|c|l|l|l|}
\hline
Problem size of ($T,\, m,\,\gamma$)  & RFA(s) & MM & EA(s)\tabularnewline
\hline
\hline
(100, 5, 10) & 0.68 & 0.115 & 1.54\tabularnewline
\hline
(150, 5, 10) & 2.22 & 0.245 & 3.07\tabularnewline
\hline
(200, 5, 10) & 4.27 & 0.434 & 6.56\tabularnewline
\hline
(250, 5, 10) & 8.03 & 0.655 & 13.10\tabularnewline
\hline
(300, 5, 10) & 13.08 & 0.950 & 19.54\tabularnewline
\hline
(350, 5, 10) & 18.54 & 1.27 & 28.08\tabularnewline
\hline
(400, 5, 10) & 25.17 & 1.70 & 38.92\tabularnewline
\hline
(450, 5, 10) & 32.48 & 2.09 & 58.56\tabularnewline
\hline
\end{tabular}}\subfloat{\centering{}%
\begin{tabular}{|c|l|l|l|}
\hline
Problem size of ($T,\, m,\,\gamma$)  & RFA(s) & MM & EA(s)\tabularnewline
\hline
\hline
(500, 5, 10) & 41.34 & 2.60 & 83.79\tabularnewline
\hline
(550, 5, 10) & 54.53 & 3.19 & 117.23\tabularnewline
\hline
(600, 5, 10) & 71.46 & 3.75 & 153.37\tabularnewline
\hline
(650, 5, 10) & 89.84 & 4.38 & 214.28\tabularnewline
\hline
(700, 5, 10) & 110.67 & 5.07 & -\tabularnewline
\hline
(750, 5, 10) & 132.80 & 6.60 & -\tabularnewline
\hline
(800, 5, 10) & 158.78 & 7.42 & -\tabularnewline
\hline
(900, 5, 10) & 217.22 & 8.35 & -\tabularnewline
\hline
\end{tabular}}
\end{table*}

\section{Conclusion }

We proposed a ratio $1-\frac{1}{e}-\epsilon$ approximation algorithm
for the $\delta$-antennae largest weight data retrieval ($\delta$ALWDR)
problem that has the same ratio as the known result but a significantly
improved time complexity of $O(2^{\frac{1}{\epsilon}}\frac{1}{\epsilon}m^{7}T^{3.5}L)$
from $O(\epsilon^{3.5}m^{\frac{3.5}{\epsilon}}T^{3.5}L)$ when $\delta=1$
\cite{lu2014data}. To our knowledge, our algorithm is the first ratio
$1-\frac{1}{e}-\epsilon$ approximation to $\delta$ALWDR for the
general case of arbitrary $\delta$. To achieve this, we first gave
a ratio $1-\frac{1}{e}$ algorithm for the $\gamma$-separated $\delta$ALWDR
($\delta$A$\gamma$LWDR) with runtime $O(m^{7}T^{3.5}L)$, under
the assumption that every data item appears at most once in each segment
of $\delta$A$\gamma$LWDR, for any input of maximum length $L$ on
$m$ channels in $T$ time slots. Then, we show that we can retain
the same ratio for $\delta$A$\gamma$LWDR without this assumption
at the cost of increased time complexity to $O(2^{\gamma}m^{7}T^{3.5}L)$.
This result immediately yields an approximation solution of similar
ratio and time complexity for $\delta$ALWDR, presenting a significant
improvement of the known time complexity of ratio $1-\frac{1}{e}-\epsilon$
approximation to the problem.

\section*{Acknowledgment}

This work is supported by Australian Research Council Discovery Project
DP150104871, \textcolor{black}{Natural Science Foundation of China
\#61300025 and} Research Initiative Grant of Sun Yat-Sen University
under Project 985. The corresponding author is Hong Shen.

\bibliographystyle{plain}
\bibliography{disjointQoS}

\section*{Appendix}

\subsection*{A Transformation from an Approximation for $\delta$A$\gamma$LWDR
to an Approximation for $\delta$ALWDR }

Let ${\cal A}$ be a ratio $\alpha$ approximation with a runtime
$t_{\delta A\gamma LWDR}$ for $\delta$A$\gamma$LWDR. Then a simple
approximation for $\delta$ALWDR with runtime $O(\frac{1}{\epsilon})\cdot t_{\delta A\gamma LWDR}$
and ratio $(\alpha-\epsilon)$ is as in Algorithm \ref{alg:Transformation-1}.

\begin{algorithm}
\textbf{Input: }A fixed $\epsilon>0$, and an instance of $\delta$ALWDR
(i.e., a set of data items to download $D=\{d_{1},\, d_{2},\,\dots,\, d_{n}\}$
with weights $w_{1},\dots,w_{n}$, together with their occurrences
in channels and time slots $t_{1},\, t_{2},\,\dots,\, t_{T}$);

\textbf{Output:} A retrieval sequence.
\begin{enumerate}
\item \textbf{For} $i=1$ to $\left\lceil \frac{1}{\epsilon}\right\rceil $
\textbf{do}

\textbf{$\quad$For} $j=1$ to $\left\lceil \epsilon T\right\rceil $
\textbf{do}

\textbf{$\quad$$\quad$}Set $t_{i+j\cdot\left\lceil \frac{1}{\epsilon}\right\rceil }$
as a vacant time slot, i.e., remove any item broadcast in $t_{i+j\cdot\left\lceil \frac{1}{\epsilon}\right\rceil }$
;

\textbf{$\quad$EndFor}

\textbf{$\quad$}Run ${\cal A}$ against the instance and obtain a
retrieval sequence $S_{i}$;

\textbf{EndFor}

\item Return $S^{*}$ with $w(S^{*})$ $=\max_{i}\{w(S_{i})\}$.
\end{enumerate}
\protect\caption{\label{alg:Transformation-1}An approximation algorithm for $\delta$ALWDR
by transformation.}
\end{algorithm}

\subsubsection*{Proof of Proposition \ref{prop:simpobsbetw-gamma}}
\begin{IEEEproof}
Let $OPT$ be an optimum solution to the original $\delta$ALWDR.
Let $\delta$ALWDR$_{i}$ be $\delta$ALWDR but with $t_{i+j\cdot\left\lceil \frac{1}{\epsilon}\right\rceil }$
vacant for each $1\leq j\leq\left\lceil \epsilon T\right\rceil $.
Assume that $OPT(i)$ is the set of items in $OPT$ being set vacant.
Then $OPT_{i}=OPT\setminus OPT(i)$ is an optimum solution to $\delta$ALWDR$_{i}$.
Let $OPT_{i^{*}}$ be the best solution among all $OPT_{i}$s. It
suffices to show $w(OPT_{i^{*}})\geq OPT\cdot(1-\epsilon)$.

From definition of $OPT_{i}$, we have
\[
w(OPT_{i})\geq w(OPT)-w(OPT(i)).
\]
Then
\[
\sum_{i=1}^{\left\lceil \frac{1}{\epsilon}\right\rceil }w(OPT_{i})=\left\lceil \frac{1}{\epsilon}\right\rceil w(OPT)-\sum_{i=1}^{\left\lceil \frac{1}{\epsilon}\right\rceil }w(OPT(i))=(\left\lceil \frac{1}{\epsilon}\right\rceil -1)w(OPT).
\]
Then
\[
w(OPT_{i^{*}})\geq OPT\cdot(\frac{\left\lceil \frac{1}{\epsilon}\right\rceil -1}{\left\lceil \frac{1}{\epsilon}\right\rceil })\geq OPT\cdot(1-\epsilon).
\]
This completes the proof.
\end{IEEEproof}
Following the above proof, we immediately have an approximation algorithm
for $\delta$ALWDR by the transformation as in Algorithm \ref{alg:Transformation-1}.

\subsection*{Performance Evaluation for $\delta$\textmd{\normalsize{}ALWDR} with
$\delta=2$}

\begin{figure}
\subfloat[$m=2$, n=100, $\gamma=2,\,\dots,\,22$.]{\begin{centering}
\includegraphics[width=0.5\columnwidth]{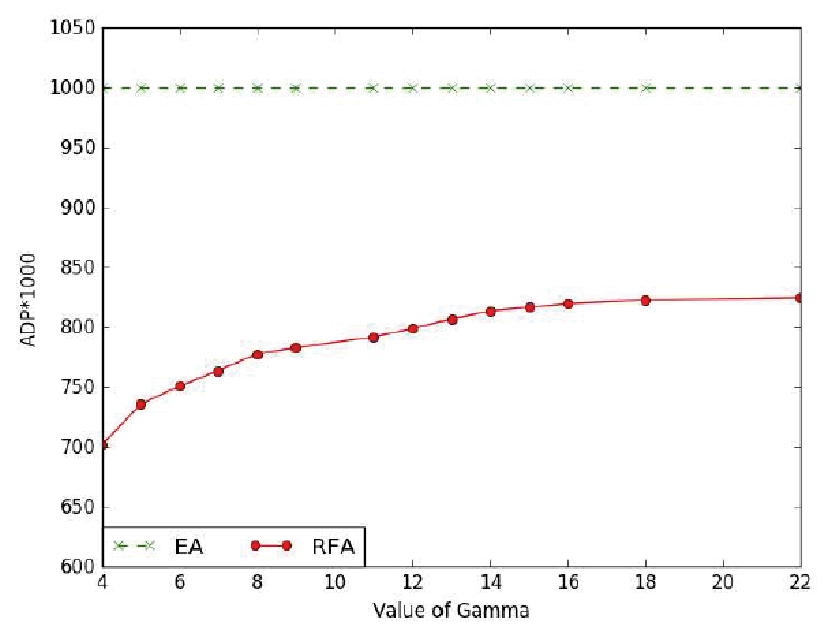}
\par\end{centering}

} \subfloat[$m=8$, n=200, $\gamma=2,\,\dots,\,22$.]{\includegraphics[width=0.5\columnwidth]{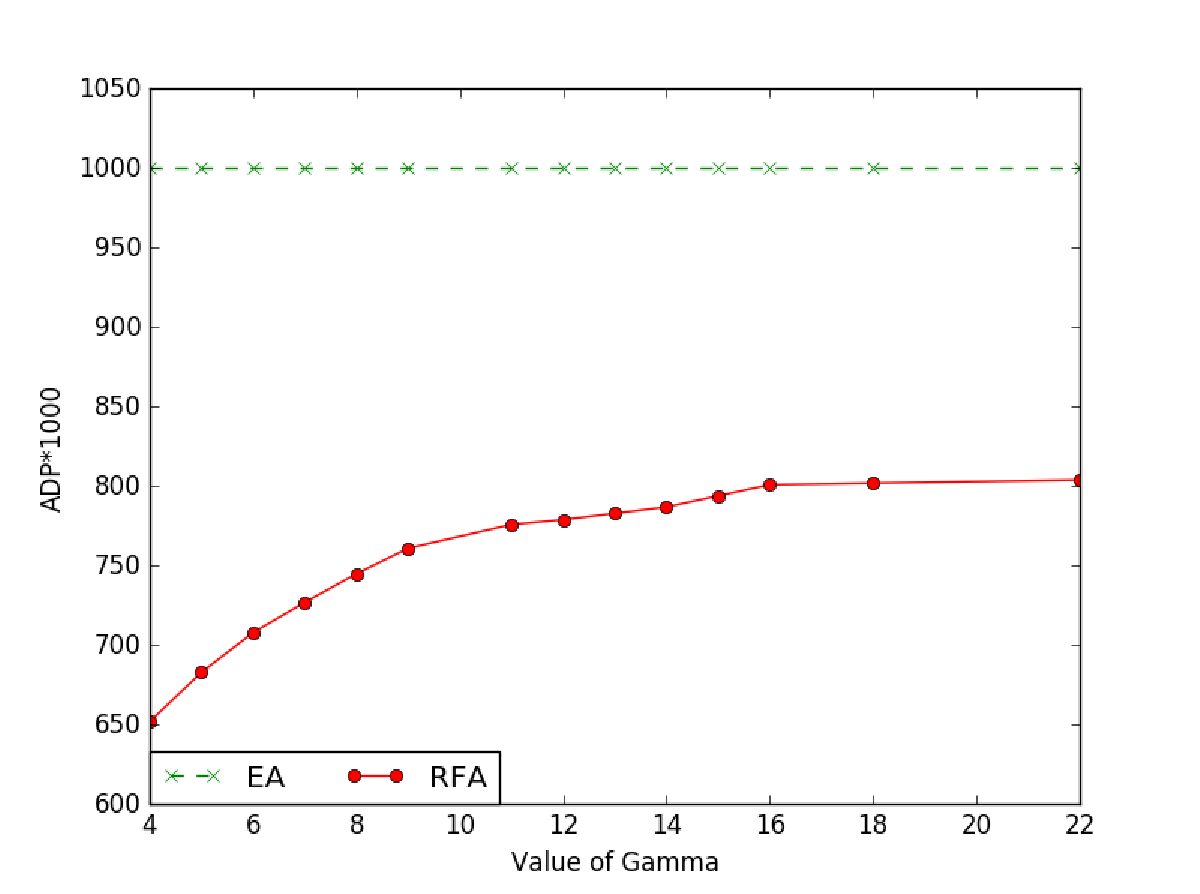}

}

\protect\caption{\label{fig:lpvsilp}$\delta$ALWDR, $T=100$, $p=150$. }
\end{figure}

For $\delta=2$, the experimental results comparing the performance
of RFA and EA are depicted in Figure \ref{fig:lpvsilp}, where $m=2$,
n=100 for Figure \ref{fig:lpvsilp} (a), $m=8$, n=200 for Figure
\ref{fig:lpvsilp} (b), and $\gamma$ grows from 2 to 22 in both figures.
The exact algorithm is to solve the integer linear programming (ILP)
formula which is LP (\ref{eq:theoriginalLP}) but with integral $x_{i,j}\in\{0,\,1\}$.
Figure \ref{fig:lpvsilp} shows that EA performs roughly 17-30 percents
better than RFA when $m=2$, and 20-35 percents when $m=8$. Since
EA always produces an optimum solution, this indicates the practical
ratio between the output solution of RFA and EA is better than our
analysis, similar to the case in Figure \ref{fig:gamma}. Besides,
also like the case in Figure \ref{fig:gamma}, when $\gamma$ grows
the performance of RFA also gets better. The increment of $\gamma$
benefits efficiently until $\gamma=14$ in Figure \ref{fig:lpvsilp}
(a) and until $\gamma=16$ in Figure \ref{fig:lpvsilp} (b).

\end{document}